\documentclass[12pt]{article}
\usepackage[utf8]{inputenc}
\usepackage[T1]{fontenc}
\usepackage[english]{babel}
\usepackage{csquotes}
\usepackage{amsmath}
\usepackage{amsfonts}
\usepackage{amsthm}
\usepackage{amssymb}
\usepackage{lipsum}
\usepackage{titling}
\usepackage{appendix}
\usepackage{float}
\usepackage{graphicx}
\usepackage[style=authoryear,backend=biber]{biblatex}
\newtheorem{hypothesis}{Hypothesis}%  meant for continuous numbers
\newcommand{\keywords}[1]{\par\noindent\textbf{Keywords:} #1}
\newcommand{\beginsupplement}{%
        \setcounter{table}{0}
        \renewcommand{\thetable}{S\arabic{table}}%
        \setcounter{figure}{0}
        \renewcommand{\thefigure}{S\arabic{figure}}%
     }
\usepackage{comment}
\usepackage{caption}
\usepackage{subcaption}
\usepackage{verbatim}
\usepackage{graphicx} % Required for inserting images
\usepackage{geometry}
\usepackage{float}
\usepackage{authblk}
\usepackage{hyperref}
\hypersetup{    
    urlcolor=blue,
    pdfpagemode=FullScreen,
    }
\usepackage[symbol]{footmisc}

\title{Divergent Characteristics of Biomedical Research across Publication Types: A Quantitative Analysis on the Aging-related Research}

\author[1,*]{Chenxing Qian}
\author[2,*]{Qingyue Guo}
\affil[1]{Department of Mathematics, University of Alabama at Birmingham,1402 10th Avenue South, University Hall, Room 4005, Birmingham, 35294, Alabama, U.S.A}
\affil[2]{Department of Molecular Biosciences, Northwestern University, 2205 Tech Drive, Evanston, 60201,  Illinois, U.S.A.}
\affil[*]{Corresponding Author}
\affil[1]{Email: cqian@uab.edu}
\affil[2]{Email: qingyueguo2022@u.northwestern.edu}
\begin{document}
\date{}

\maketitle

\begin{abstract}
This paper investigates differences in characteristics across publication types for aging-related genetic research. We utilized bibliometric data for five model species retrieved from authoritative databases including PubMed. Publications are classified into types according to PubMed. Results indicate substantial divergence across publication types in attention paid to aging-related research, scopes of studied genes, and topical preferences. For instance, comparative studies and meta-analyses show a greater focus on aging than validation studies. Reviews concentrate more on cell biology while clinical studies emphasize translational topics. Publication types also manifest variations in highly studied genes, like APOE for reviews versus GH1 for clinical studies. Despite differences, top genes like insulin are universally emphasized. Publication types demonstrate similar levels of imbalance in research efforts to genes. Differences also exist in bibliometrics like authorship numbers, citation counts, etc. Publication types show distinct preferences for journals of certain topical specialties and scope of readership. Overall, findings showcase distinct characteristics of publication types in studying aging-related genetics, owing to their unique nature and objectives. This study is the first endeavor to systematically depict the inherent structure of a biomedical research field from the perspective of publication types and provides insights into knowledge production and evaluation patterns across biomedical communities.
\end{abstract}

% \section{Heading 1}
%    So, for example \parencite[]t{hasselmo95} investigated that\dots
% \begin{thebibliography}{99} 
%   \bibitem[Alexander et al.(1995)]{Alexander95} Alexander, J.A.\ \& Mozer,  M.C.\ (1995) Template-based algorithms
%    for connectionist rule extraction. In G.\ Tesauro, D.S.\ Touretzky and
%    T.K.\ Leen (eds.), {\itshape Advances in Neural Information Processing
%    Systems 7}, pp.\ 609--616. Cambridge, MA: MIT Press.
%   \bibitem[Hasselmo et al.(1995)]{hasselmo95} Hasselmo, M.E., Schnell, E.\ \& Barkai, E.\ (1995) Dynamics of
%     learning and recall at excitatory recurrent synapses and cholinergic
%     modulation in rat hippocampal region CA3. {\itshape Journal of
%     Neuroscience} {\bfseries 15}(7):5249-5262.
% \end{thebibliography}
\section{Introduction}\label{sec1}
Research publications serve as a means of conveying knowledge, findings, and novel topics in the field of biomedical sciences. Researchers across different communities and cultures rely on prior literature to develop and build upon existing facts, knowledge, and methodologies to advance scientific discoveries. It is widely acknowledged that the types of publications used to embody the fundamentals, motivations, and objectives of upcoming studies vary across various biomedical communities. 

Each publication type in biomedical sciences has its unique nature, purpose, and evaluation criteria \parencite[]{wcg2008basic}. The variations in the use of scientometric characteristics (such as citation counts, \textit{etc.}), the collaboration forms, and the roles played in scientific communication come with the different publication types \parencite[]{bourke1996publication,andersen2023field}. Besides, there might be rudimentary differences in the objects studied, which are associated with the epistemological and ontological assumptions across publication types,  as \parencite[]{andersen2023field} indicated for the field-level differences. Exploring different publication types can thus shed light on the production patterns of various biomedical knowledge from a fresh perspective. Moreover, the dissemination of knowledge through different publication types may hold different levels of value \parencite[]{bourke1996publication}. Therefore, it is crucial to investigate the scopes and overlaps of topics covered by varying publication types to redistribute research efforts toward more valuable areas. With the increasing volume of biomedical publications, it's becoming challenging for small groups and individual researchers to collect related studies from all aspects \parencite[]{hunter2006biomedical}. Thus, further investigations on publication types can guide data mining efforts to improve their efficiency and accuracy. 

It is becoming increasingly common to see research revealing the scientometric trends of a field (from large-scale as physics and small-scale neuroimaging) leveraging the voluminous available bibliographic metadata. However, to the best of our knowledge, most existent literature only focused on the divergences of characteristics of a limited choice of biomedical publication types, regarding the evidence levels of medical reporting, on a small range of studies, including some specific clinical subdomains \parencite[]{patsopoulos2005,mays2012,rosenkrantz2016most,yeung2020most,fontelo2018}. Additionally, no published literature has focused on the scopes of genes, which are at the core of biological research due to their indispensable ontological significance, studied by different publication types.

This article provides a comparative overview of differences regarding most common publication types in biomedical (especially aging-related) research, including ordinary research articles, comparative studies, clinical studies, reviews, meta-analyses, systematic reviews, evaluation studies, validation studies, \textit{etc.} In this paper, we hypothesized that the characteristics of biomedical publications will show the differences regarding the scopes and preferences on topics and genes in various aspects, across the types of publications with a focus on aging-related research. We used the systems categorizing publication types defined by NLM (National Library of Medicine) as proxies for the forms of publication universal for biomedical studies (Section \ref{data_retrieval}). We performed a quantitative and exploratory analysis of the bibliometric data retrieved from PubMed (Section \ref{data_retrieval}). 

There are several key reasons for focusing on aging-related genetic research. Firstly, aging is a universally recognized and inevitable biological process that has captured the attention of scientists for decades \parencite[]{mccay1935effect,lopez2013hallmarks}. As a result, rich biology has been accumulated in this field for humans and other species. Secondly, aging is closely linked to many other biological processes \parencite[]{lopez2013hallmarks,jones2015dna,kitao1999mutations,kennedy2014geroscience}, and the development of other scientific fields has contributed to the advancement of aging research. For instance, the increased understanding of genetics has led to the identification of links between the aging process and specific genes \parencite[]{kenyon2010genetics}. Aging-related studies encompass various types of publications as they cover a broad scope of medical practices. In addition, we limited this analysis to genetic research as a way to minimize research dominated by the spirits of fields other than biomedical sciences.

We hope that the presenting analysis can provide insights into advancing the understanding of aging-related research and improving the current knowledge production, dissemination, and communication systems in biomedical sciences.

\section{Literature review}
\subsection{Explaining the various forms of publication in biomedical sciences}
The field of biomedical sciences, after the advent of large-scale bioinformatics \parencite[]{barabasi2011network,xia2017bioinformatics} and \textit{in silico} computational techniques \parencite[]{gentleman2005bioinformatics,shortliffe2014biomedical}, encompass a broad spectrum of topics, ranging from elucidating molecular mechanisms, explaining unknown genotype-phenotype correlations, to enhancing the applicability of medical practices. To date, this field has gathered resources, ideas, and scholars from diverse communities and organizations, resulting in significant research output \parencite[]{conte2017globalization,ismail2012bibliometrics,beaudevin2016diversion}. The biomedical sciences today are a grand assembly of idiosyncratic cultures, philosophies, and beliefs.

As biomedical sciences increasingly adopt transdisciplinary knowledge production patterns \parencite[]{soofi2018mode}, it is important to recognize that complexity is not only a result of interdisciplinary components but also intrinsic to biology \textit{per se}, which is at the core of biomedical science \parencite[]{hacking1983representing}. Unlike the "normal science" of Kuhn, research patterns in biology are complex and shaped by the ways experimental phenomena are produced and maintained \parencite[]{boem2016}. Boundaries between studies, such as confirmatory/exploratory and theory-informed/theory-driven, are hard to establish \parencite[]{waters2007,omalley2007,elliott2007}. \parencite[]{rheinberger1997} summarized the processes of exploring liminal biological knowledge as "the modern biologist is bound to divide his world into fragments in which parameters can be defined, quantities measured, qualities identified." Thus, distinct cultures across research communities might involve varying topics and objects dependent research patterns, which may shape the writing styles and formats adopted by researchers in a particular field. Distinct motivations, objectives, and backgrounds held by research groups lead to utterly different epistemic cultures \parencite[]{knorr1999,ratti2021}. The different systems of generating, validating, and developing hypotheses among communities might require various types of evidence and data, and induce various conclusions even given the identical results \parencite[]{rheinberger1997,waters2007,franklin2005,ratti2021}. Moreover, the “halo effects” of previous influential studies in one domain would shape the organization of ideas and representation of evidence of upcoming publications \parencite[][pp.~40]{hofmann2010, schimel2012}. Lastly, the reviewers during the research evaluation processes may expect certain regimes of well-structured organizations of manuscripts as the past acknowledged publications.

The divarication of research patterns can be even amplified during the publication processes, which involve the representation, dissemination, evaluation, and selection of knowledge and findings. With the breakdown of disciplinary boundaries, the exponential increase of scientific publications \parencite[]{hunter2006biomedical,white2019}, and the increasingly inevitable specializing trends across biomedical subdomains, individuals, organizations, together with the biomedical communities, had worked on establishing platforms and systems integrating all structured achievements, findings, and knowledge in all active fields of biomedical and life sciences \parencite[]{terry2014global}. As a result, thousands of new biomedical journals emerged for the purposes of recording and classifying the new knowledge recognized and generated by newly admitted biomedical subdomains \parencite[]{fiala2017emerging,gasparyan2009medium,yuri2011mean}. 

Notably, as the ways of research evaluation by most journals, adapted forms of peer review forward the establishment of divergent publishing standards. \parencite[]{maisonneuve2000peer,lipworth2011journal}. The reviewers involved are “stakeholders” holding pluralistic epistemic (and nonepistemic) values considering different facets of a system \parencite[][pp.~55]{hackett1997}. Biological sciences comprise \textit{sui generis} decentralized systems of cultures upon which dispersed research efforts are paid \parencite[][pp.~82]{knorr1999}. The different opinions, values, and even chasms in the understanding of critical problems across biomedical subdomains constituted of researchers holding heterogeneous backgrounds and beliefs, might induce and entrench discrepancies in evaluation and valuation standards. Over time, divergent standards for organizing studies across subdomains may be reflected in academic writings, resulting in scientific publications produced under distinct patterns and presented in various forms. In addition, reviewers tend to regularize comments to an acceptable cliché which are often confined to technical issues such as formative standards and some “compulsory” modifications, including quantitative analysis and experimental results \parencite[]{ploegh2011}, as the results of the compromission within the epistemic authorities \parencite[][pp.~54]{nightingale2007peer,abrams1991predictive,hackett1997}, causing the published manuscripts in one subdomain to be “averaged” or homogenized, and leading to the peer review process being a double-edged sword, which could be detrimental to the emergence and dissemination of innovations \parencite[]{de1963scientific,osterloh2020avoid,crane1967gatekeepers}. 

\subsection{Previous literature on publication types in biomedical sciences}
Based on the argument above, the varying consensus on accepting and publishing new knowledge across dispersed biomedical communities contributes to the development of subdomain-dependent canonical “study forms” (in which research patterns and representing forms are both implicated). For example, the research on the functions of molecular components in a pathway and the contributors of diseases is often related to the comparative genomic analysis across conditions \parencite[]{sato2004functional,alfoldi2013comparative}. The clinical investigations on the effectiveness and efficacy of treatments are reported via highly structured and clearly object-oriented formats requiring details on statistical powers and study designs \parencite[]{faggion2012guidelines,lin2012guidelines,zeng2015methodological}. Narrative reviews and systematic reviews (including meta-analyses) are differently accepted and valued by a varying scope of biomedical communities \parencite[]{williams2017meta,stavrou2014archibald}. To comprehensively integrate and archive biomedical information, previous studies have classified biomedical research into publication types, including case reports, systematic reviews, and comparative studies, \textit{etc} \parencite[]{pubtypes,wcg2008basic}. This approach provides a standard framework for categorizing research paradigms and publishing standards. Each type of publication is unique in its nature, serves a distinct purpose, and is evaluated using specific criteria.

Previous studies have investigated the variations in contextual and bibliometric characteristics across different types of publications in the biomedical sciences \parencite[]{bourke1996publication,beattie2004publishing,egbert2015publication,puuska2010effects}. However, these studies have only categorized publications into three main types: journal articles, books, and conference proceedings, which is an inadequate classification, particularly for biomedical sciences where "journal articles" encompass a heterogeneous set of studies, including evaluation studies, comparative studies, and reviews. Similar studies on bibliometric characteristics specific to biomedical sciences \parencite[]{patsopoulos2005,mays2012,rosenkrantz2016most,yeung2020most,fontelo2018} have focused on a limited scope of publication types, such as case reports, cohort studies, population studies, clinical trials, meta-analyses, and systematic reviews, associated with clinical and practical biomedical sciences, for the dissemination of medical knowledge on different evidence levels. It is worth highlighting that the bibliometric characteristics and trends of publications that fall under broader scopes of publication types and subdomains, such as genetic research in the form of comparative studies, evaluation studies, and reviews, and those related to biomedical subdomains, have rarely been studied. Therefore, there is an urgent necessity for further investigations into the differences in characteristics of publications across various types, because of the plausible correlations to the cultures and standards of different biomedical communities, as aforementioned. Furthermore, the awareness of these differences might benefit in informing the development of more effective acceptance and dissemination strategies for novel findings and facilitating more comprehensive and accurate data mining and meta-analyses.
\subsection{The landscape of aging-related research}
Aging, widely recognized as the progressive functional decline (and loss) that most organisms are involved in, has long been a subject of curiosity and investigation throughout history \parencite[]{lopez2013hallmarks}. The success of isolating the longevity mutants of \textit{C. elegans} \parencite[]{klass1983method} ignited the establishment of new research paradigms of aging by the modern technologies and methodologies of manipulating molecular and cellular bases. Aging research has now become an active line encompassing almost all subdomains in biomedical sciences, including genetic, epigenetic, neurological, biochemical, molecular, and cellular mechanisms of aging, as well as their influence on aging-related diseases \parencite[]{lopez2013hallmarks,jones2015dna,kitao1999mutations,kennedy2014geroscience}. Aging research holds immense biological and medical significance for numerous reasons. The investigations into aging will shed light on the biological fundamentals such as cellular senescence, telomere attritions, and mitochondrial function \parencite[]{lopez2013hallmarks,gil2016out}. The understanding of these processes can thus inform the development of preventive medicine, therapies, and treatments in broader scopes of medical research, such as Alzheimer's disease, Parkinson's disease, cardiovascular diseases, cancer biology, and regenerative medicine \parencite[]{gu2019healthy}. To date, many disciplines emerged as specializations of aging-related research on different topics. The importance of aging-related research is also reflected in the increasing annual entries of publications in aging-related subdomains \parencite[]{gu2019healthy,othman2022profiling,jiang2023knowledge} horizontal to the increasing biomedical publications. 

\section{Hypotheses, study designs, and methodologies}
\subsection{Hypotheses}\label{hypotheses}
The fundamental premise of this paper is that publications represent different study designs and thus reflect varying characteristics. In this section, we develop several hypotheses based on previous research and discussions.

Here we provided the first hypothesis:
\begin{hypothesis}
\normalfont For one specific biomedical subdomain, researchers will show different levels of preferences on the types of publications, according to the distinct purposes, methodologies, and criteria in this subdomain.
\end{hypothesis}

We find the first hypothesis to be reasonable, supported by relevant literature. For instance, \parencite[]{austad2009comparative} contended that comparative analyses performed on biosystems of different sizes and lifespan scales could enhance the understanding of aging processes. Meta-analyses and systematic reviews, with evidence synthesized from multiple studies, possess higher levels of confidence due to stringent selection criteria \parencite[]{sackett1997evidence,petrisor2006grading}. The academic pressures \parencite[]{yang2013meta}, availability of clinical data on aging processes, stronger evidence, and greater chances of obtaining citations \parencite[]{patsopoulos2005}, and the critical healthcare significance, can prompt the production of aging-related syntheses and reviews with high levels of evidence. 

We can further develop the second hypothesis, based on these claims, that the topics and genetic entities involved in various publication types differ due to varying motivations, objectives, epistemic cultures, experimentation and evaluation procedures, and even different scopes of previous studies to be referenced. For instance, comparative studies tend to focus on functional hotspots of aging processes, while systematic reviews and meta-analyses may report more on genetic entities with clinical significance.

\begin{hypothesis}
    \normalfont As the objectives, nature, and evaluation of different publication types vary, there are discrepancies in the scopes of genes (and the corresponding products) and the biomedical topics across different publication types. 
    \label{hypothesis_2}
\end{hypothesis}

Previous studies have highlighted the disproportionate attention given to a small subset of human protein-coding genes, which is partly due to historical and technical factors rather than their biological significance \parencite[]{stoeger2018large,haynes2018gene,stoeger2022characteristics}. We contend that this imbalance persists even within a single publication type. Furthermore, we propose that it may also extend to other biomedical topics, as different publication types have distinct formats and styles that facilitate reporting new findings from diverse subdomains. We further posit that the specific subset of genes that attracts the most attention varies across publication types due to their distinct nature, objectives, and evaluation criteria, as we previously mentioned.
\begin{hypothesis}
    \normalfont Focusing on each publication type, the research attention to genes and biomedical topics is unbalanced, as embodied in the publication and citation counts. The genes receiving the most research attention vary across publication types, corresponding to the nature, objectives, and evaluations of different publication types. 
    \label{hypothesis3}
\end{hypothesis}

There is adequate literature examining the bibliometric characteristics specific to biomedical publications. For example, \parencite[]{leydesdorff2013citation} and \parencite[]{leydesdorff2012bibliometric} investigated the differences in citation behaviors across medical subject headings (MeSH). \parencite[]{burrows2011trends} investigated the representation of authorship information in biomedical publications. \parencite[]{stoeger2022characteristics} explored publishing contexts, authorship, funding, and citation counts of biomedical publications focused on early-stage or upcoming genetic research. \parencite[]{fontelo2018} reported temporal trends in the annual entry of biomedical publications across different publication types. Similarly, studies specific to various medical subdomains have also been conducted, such as \parencite[]{yeung2020most} and \parencite[]{rosenkrantz2016most}. Furthermore, \parencite[]{patsopoulos2005} uncovered differences in citation impact among various publication types in biomedical research. Hence, it is reasonable to expect differences in bibliometric behaviors across different publication types in the biomedical field.

\begin{hypothesis}
    \normalfont The bibliometric characteristics of publications are divergent based on the publication types, including publication and citation counts, authorship information, funding history and resources, and journals appreciated.
    \label{hypothesis4}
\end{hypothesis}

%\section{Data Extraction and Preprocessing}\label{section2}
\subsection{Data retrieval}\label{data_retrieval}

The bibliometric data were retrieved from PubMed. We chose 5 representative species intensively studied both inside and outside aging-related fields (human, mouse, rat, fly, and \textit{C.elegans}) for further analysis. We restricted our analysis to publications directly related to genetic research and separated the aging-related and nonaging-related publications by the following steps. We used curated data from PubTator \parencite[]{wei2019pubtator} to filter publication records only related to genetic research by including publications mentioning the symbols (or synonyms) of genes in titles and abstracts at least one time. As the names, symbols, and NCBI IDs are different for genes from different species, we obtained and separated the biomedical genetic publications on the 5 representative species. 

Thereafter, we further classified the publications as "aging-related" by including the publications tagged by MeSH terms \parencite[]{dhammi2014medical} about aging ("Aging", “Cellular Senescence”, “Erythrocyte Aging”, “Senescence-Associated Secretory Phenotype”, “Telomere Shortening”, “Cognitive Aging”, “Immunosenescence”, “Longevity”). The rest of the genetic publications were classified as "nonaging-related". As the information on publication types of each retrieved publication was included in PubMed, we further categorized each "aging-related" and "nonaging-related" publication using a system of publication types modified from the original NLM-defined publication type classification rules (Table \ref{table1}), to assemble analogous publication types and classify publication types without results and discussions meeting certain specific biomedical criteria. Detailed procedures on data retrieval were recorded in Section \ref{appendix_data_retrieval}. 

Additionally, to provide a more accurate portrayal of the characteristics of each publication type, we also counted the citation counts (the total number of being cited) of "aging-related" and "nonaging-related" publications for each type, using the iCite databases \parencite[]{hutchins2019nih}. The specific procedures for accessing and collecting these citation records are outlined in Section \ref{appendix_citation}.
% We restricted our analysis to publications in the biomedical field. Publication information including PubMed ID, publication type, published date and MeSH \parencite[]{dhammi2014medical} terms tagged was extracted from PubMed database. Gene information including NCBI ID, symbol, description, and taxonomy ID was obtained from NCBI. We excluded publications published later than 2020 due to insufficient recording and editing. We also collected the citation information for each publication from PubMed. For each individual publication, the PubMed identifiers (PubMed id) of upcoming publications which cited it was retrieved from iCite database \parencite[]{hutchins2019nih}.

% In order to narrow our focus on scientific publications in the biomedical sciences, we chose the following publication types for investigations: “Review”, ”Introductory Journal Article”, ”Comparative Study”, ”Historical Article”, ”Evaluation Study”, “Validation Study”, “Twin Study”, “Classical Article” and 4 classes (Table \ref{table1}) of publications which were merged from some remaining individual types because of similar research objectives and methodologies. 

% We chose 5 representative species intensively studied both inside and outside aging-related fields (human, mouse, rat, fly, and C.elegans) for analysis. After sorting the data, 3863561 publications on human genes, 847282 publications on mouse genes, 550865 publications on rat genes, 57346 publications on fly genes, and 12471 publications on \textit{C.elegan}s genes were retained. 
\begin{table}[ht]

\caption{Mapping the merged publication categories in this paper to the original categories by PubMed. The merged categories contain studies sharing similar research and bibliometric features. }
\centering
\begin{tabular}{rll}
  \hline
  \hline
  \multicolumn{1}{l|}{Merged Categories} & Original Categories \\ 
  \hline
  \hline
 \multicolumn{1}{l|}{Research Article} & Journal Article;  \\
 \multicolumn{1}{l|}{}& Corrected and Republished Article \\
 \multicolumn{1}{l|}{}&(excluding the publications that belong to other types) \\ 
 \hline
   \multicolumn{1}{l|}{Clinical Study} & Pragmatic Clinical Trial;\\
   \multicolumn{1}{l|}{}& Observational Study, Veterinary; \\
   \multicolumn{1}{l|}{}& Clinical Conference; Clinical Study;\\
   \multicolumn{1}{l|}{}& Clinical Trial, Veterinary; \\
   \multicolumn{1}{l|}{}& Clinical Trial Protocol; Clinical Trial, Phase I; \\
   \multicolumn{1}{l|}{}& Clinical Trial, Phase II; Clinical Trial, Phase III; \\ 
   \multicolumn{1}{l|}{}& Clinical Trial; Randomized Controlled Trial; \\
   \multicolumn{1}{l|}{}& Case Reports; Controlled Clinical Trial; \\
   \multicolumn{1}{l|}{}& Observational Study;  \\ 
   \hline
   \multicolumn{1}{l|}{Meta-Analysis, Systematic Review} & Meta-Analysis; Systematic Review \\ 
   \hline
   \multicolumn{1}{l|}{Comment, Letter, News} & Comment; News; Letter \\ 
   \hline
   \hline
\end{tabular}
\label{table1}

\end{table}

\section{Results}\label{sec2}
\subsection{Divergent levels of preferences to aging-related research for various publication types}
\begin{figure}[H]
    \centering
    \includegraphics[width=1\columnwidth]{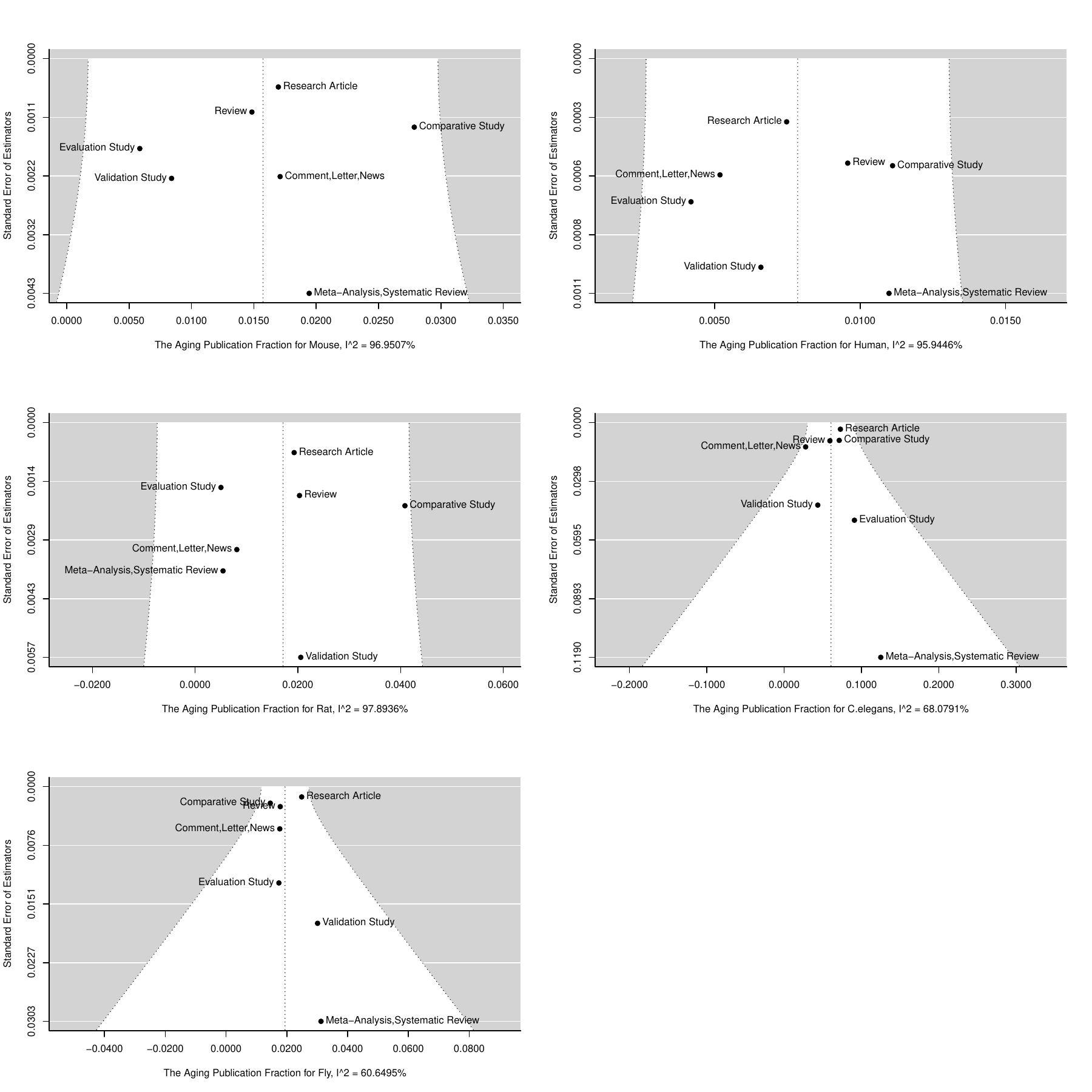}
    \caption{We displayed the fraction of attention by seven important publication types, including research articles, reviews, comparative studies, meta-analyses \& systematic reviews, evaluation studies, validation studies, and comments \& letters \& news, across five species in the funnel plots. The dashed lines in the middle of the subfigures represent the average attention to aging for one species. The large widths of the white “funnels” suggest a strong heterogeneity of attention among the seven publication types. The $I^2$ quantity, which represents the level of heterogeneity, is appended to the titles of subfigures. }
    \label{figure1}
\end{figure}

To demonstrate the diverging attention paid to aging-related research relative to nonaging-related research by different publication types, we employed the RE model (Section \ref{section3.1}) using the publication and citation fractions (Section \ref{appendix_citation}) to show the variability of attention to the aging-related field across publication types and five frequently studied species: Human, Mouse, Rat, \textit{C.elegans} and fly (Figure \ref{figure1}, Figure \ref{figureS1} and Figure \ref{figureS2}). 

%As shown in Figure \ref{figure1} and Figure \ref{figureS2}, the average attention to aging-related research varies among species. Aging-related research on \textit{C.elegans} receives the highest attention, which is 6.07\% in  publication fraction, among the five species. Human, Mouse, Rat and Fly receive attention ranging from 1\% to 2\% in publication fractions, where attention to Human and Fly is the lowest and highest respectively. 

As shown in Figure \ref{figure1} and Figure \ref{figureS2}, the average attention to aging-related research varies among publication types in all 5 species. The large widths of the funnels especially in subfigures of mouse, human, and rat indicate an apparent divergence of attention to aging among selected publication types. The statistical diagnostics indicated the strong divergence of attention, quantified as the heterogeneity in the RE model, by various publication types (Table \ref{tableS1}). Among the five species, \textit{C.elegans} and Fly showed more consistency in the attention across publication types compared to Human, Mouse and Rat. Considering that \textit{C.elegans} and Fly are important model animals for aging, they may have less versatile functions in biomedical research. The divergence of publication types, which stems from the complexity of biology, is much less manifested in \textit{C.elegans} and Fly.

Despite the different levels of heterogeneity, there are similar patterns between species. From Figure \ref{figure1}, we can observe that the attentions by research articles and reviews are proximate to the average attention considering all publication types to aging-related research for these species. In general, evaluation studies, clinical studies, and validation studies showed less attention to aging-related research, while the attention by meta-analyses \& systematic reviews, and comparative studies exceed the average attention in the five species (Figure \ref{figure1} and Figure \ref{figureS1}). In sum, Figure \ref{figure1} and Figure \ref{figureS1} visually validated Hypothesis 1 and suggested considerable preferences on aging-related studies by specific comparative studies, clinical studies and validation studies.

\begin{figure}[H]
    \centering
    \includegraphics[width=1\columnwidth]{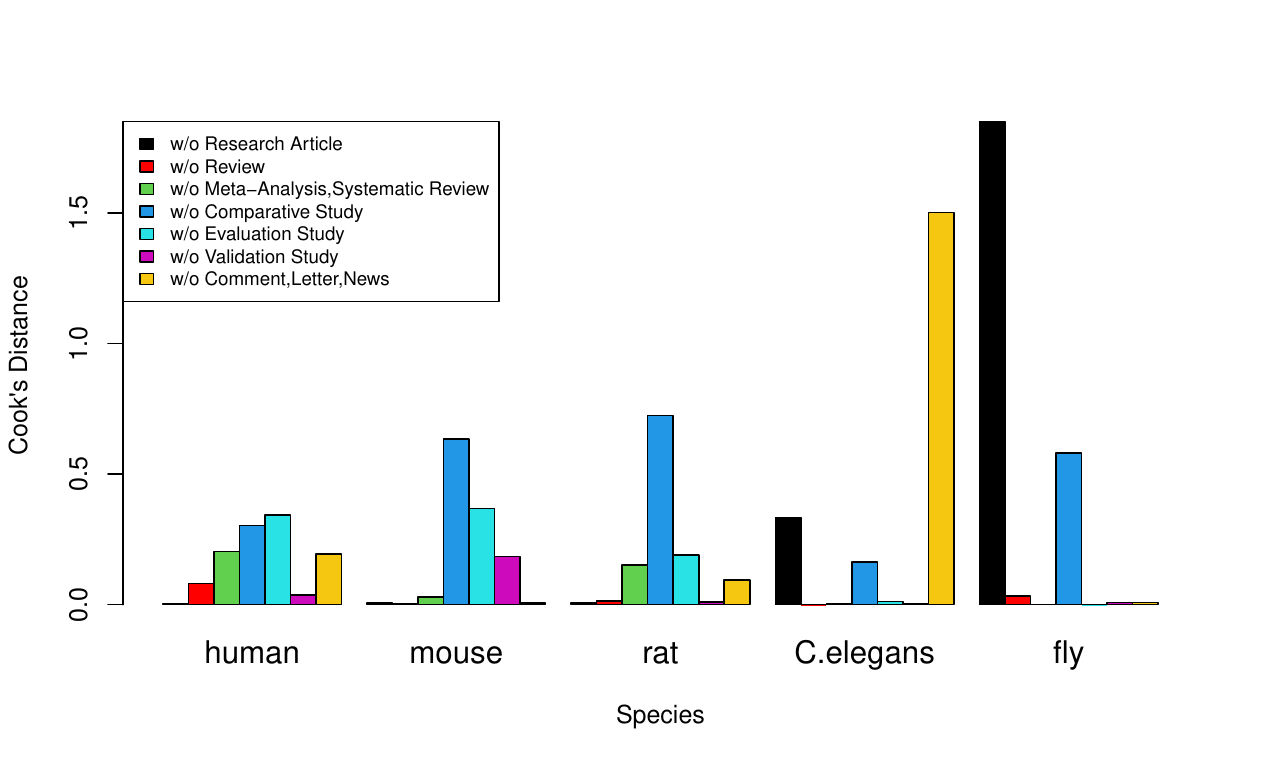}
    \caption{We displayed the Cook’s distance, which measures the amount of heterogeneity contributed by each publication type in the RE model, for the five species by removing each publication type out repeatedly.
}
    \label{figure2}
\end{figure}

Among the seven publication types, we calculated Cook’s distance, a measure of heterogeneity contributed by each publication type, in the five species by removing one publication at each time (Figure \ref{figure2}).  For Human, Mouse and Rat, comparative studies and evaluation studies are major contributors to attention divergence, suggesting distinct research patterns and focus guided by academic styles. For \textit{C.elegans} and Fly, the research articles showed more intensive attention to aging-related research compared to other publication types. 

\subsection{Different scopes of aging-related topics and genes were investigated by various publication types}
There exist huge variations in the genes studied and preferences on genes among various publication types. For example, clinical studies and mechanism studies show contrasting interests in how and where the genes function.  Here we quantitatively demonstrated the variation of research attention on genes by several approaches. 

\begin{figure}[H]
\begin{subfigure}{.5\textwidth}
  \centering
  % include first image
  \includegraphics[width=1\linewidth]{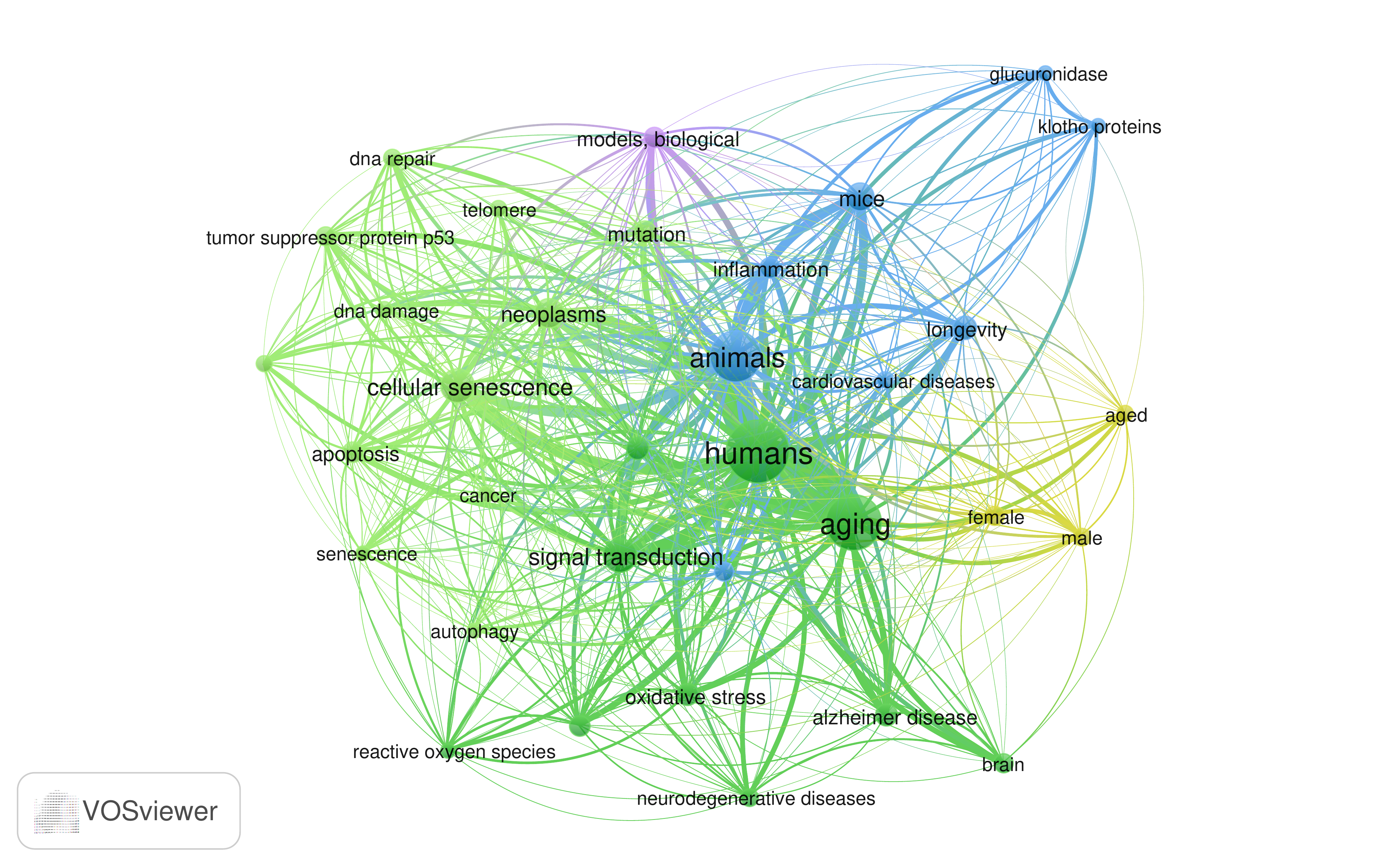}  
  \caption{Reviews}
  \label{fig4:sub-first}
\end{subfigure}
\begin{subfigure}{.5\textwidth}
  \centering
  % include second image
  \includegraphics[width=1\linewidth]{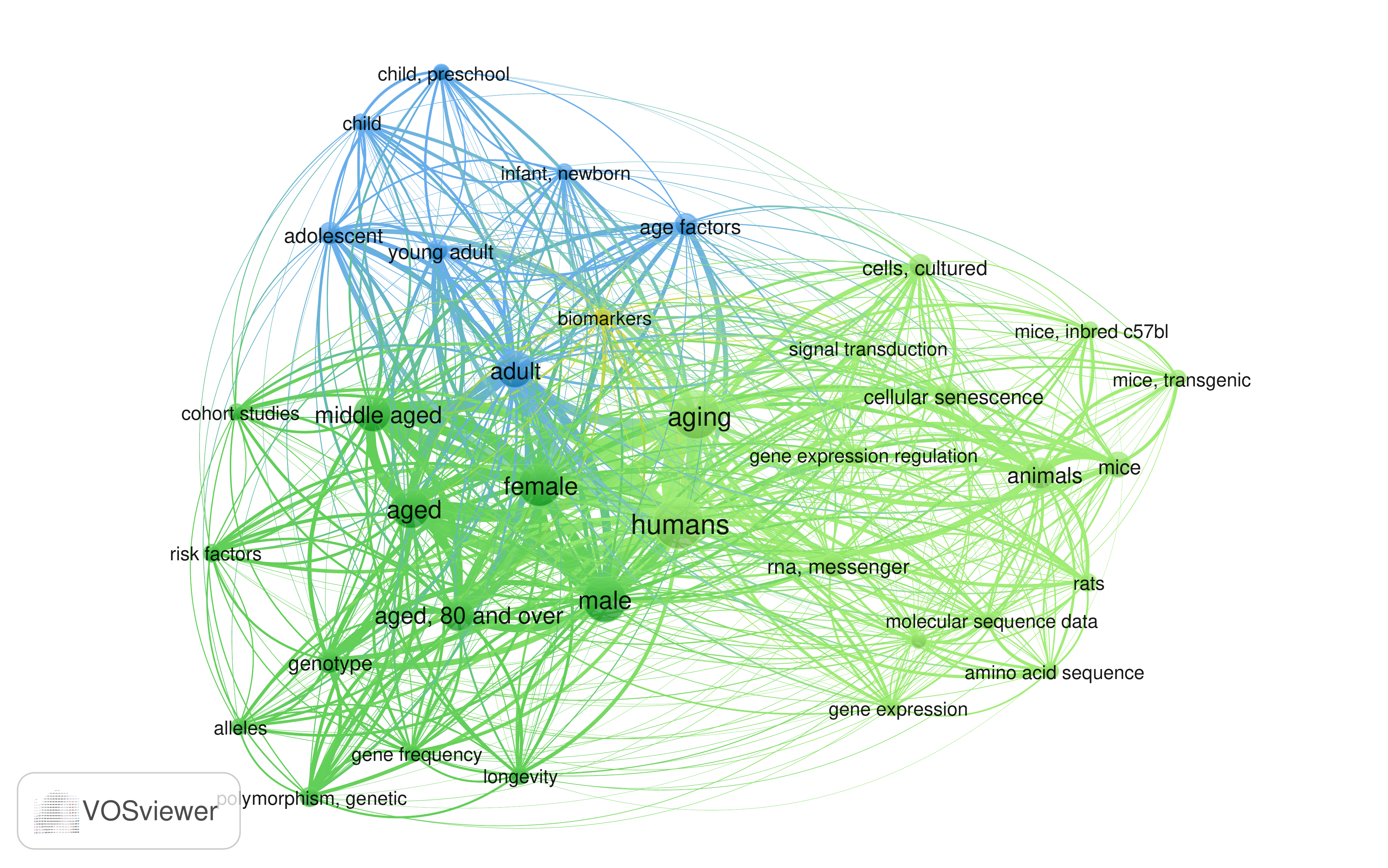}  
  \caption{Comparative studies}
  \label{fig4:sub-second}
\end{subfigure}
\\
\begin{subfigure}{.5\textwidth}
  \centering
  % include second image
  \includegraphics[width=1\linewidth]{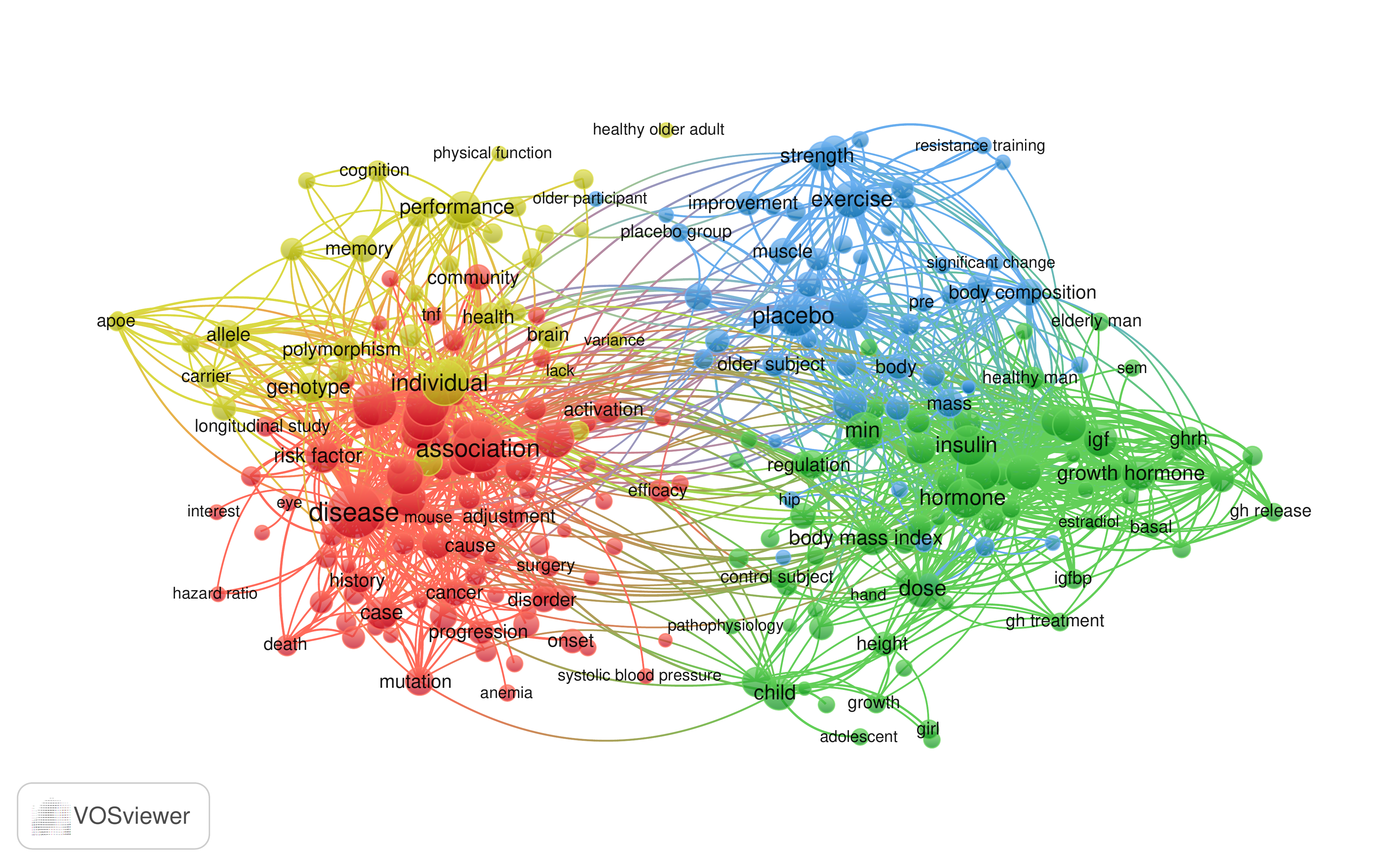}  
  \caption{Clinical studies}
  \label{fig4:sub-third}
\end{subfigure}
\begin{subfigure}{.5\textwidth}
  \centering
  % include second image
  \includegraphics[width=1\linewidth]{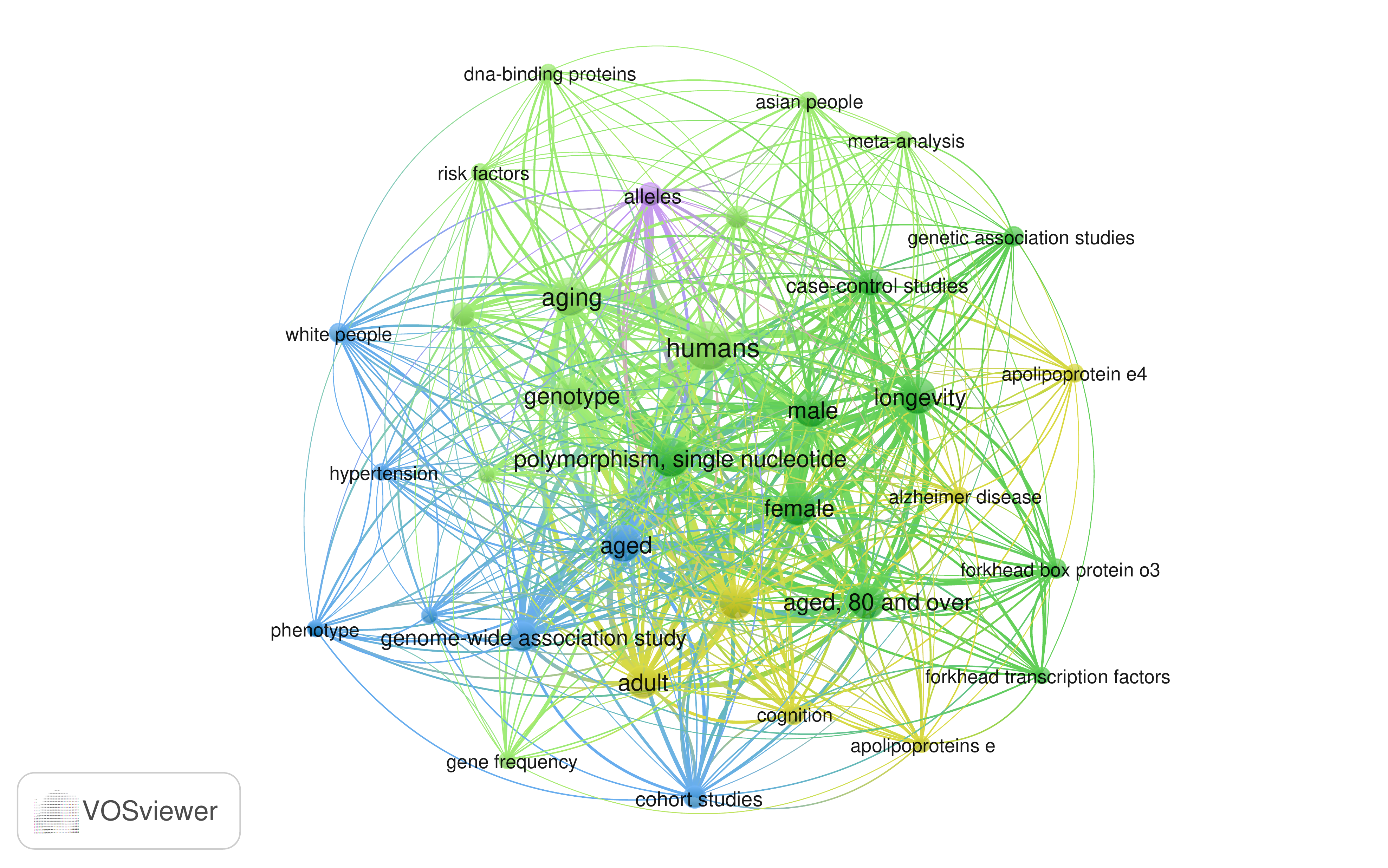}  
  \caption{Meta-analyses and systematic reviews}
  \label{fig4:sub-forth}
\end{subfigure}
\caption{The co-occurrence maps (by VOSviewer) of research topics based on the texts of titles and abstracts of literature from four publication types. Each color refers to one cluster based on topical relevance.}
\label{figure4}
\end{figure}

In Section \ref{hypotheses} we proposed the hypothesis on the inconsistent scopes of research objects (Hypothesis \ref{hypothesis_2}) across publication types. Figure \ref{figure4} partially validated the distributions and connections of research topics under four publication forms by clustering the MeSH terms of publication data retrieved from NCBI (Section \ref{data_retrieval}) via VosViewer \parencite[]{van2010software,van2011vosviewer}. Reviews, comparative studies, clinical studies, and meta-analyses \& systematic reviews show distinct emphasis on ranging from medical care to cell biology. In contrast to reviews that majorly focus on the signal transduction on cellular senescence, the comparative studies centered on three more detached subfields: genotypic studies on aging, gene expression on cellular senescence and human development biology regarding the age factors during different stages (Figure \ref{fig4:sub-first} and \ref{fig4:sub-second}). As a major publication type, the clinical studies even demonstrated a more bifurcated trend on topics (Figure \ref{fig4:sub-third}): one class (red and yellow) consists of studies on clinical progressions of diseases including genotypic and polymorphism research, the opposite class is composed by studies on the regulations and impacts of hormones, as well as related kinesiological studies. Besides genotypic and longevity studies (which are also conveyed via the proceeding three publication types), meta-analyses and systematic reviews, usually utilized as approaches synthesizing and integrating widely-sourced data and conclusion, naturally emphasized genome-wide association studies (GWAS), cognition decline and hypertension (Figure \ref{fig4:sub-forth}).

\begin{figure}[H]
    \centering
    \includegraphics[width=1.1\columnwidth]{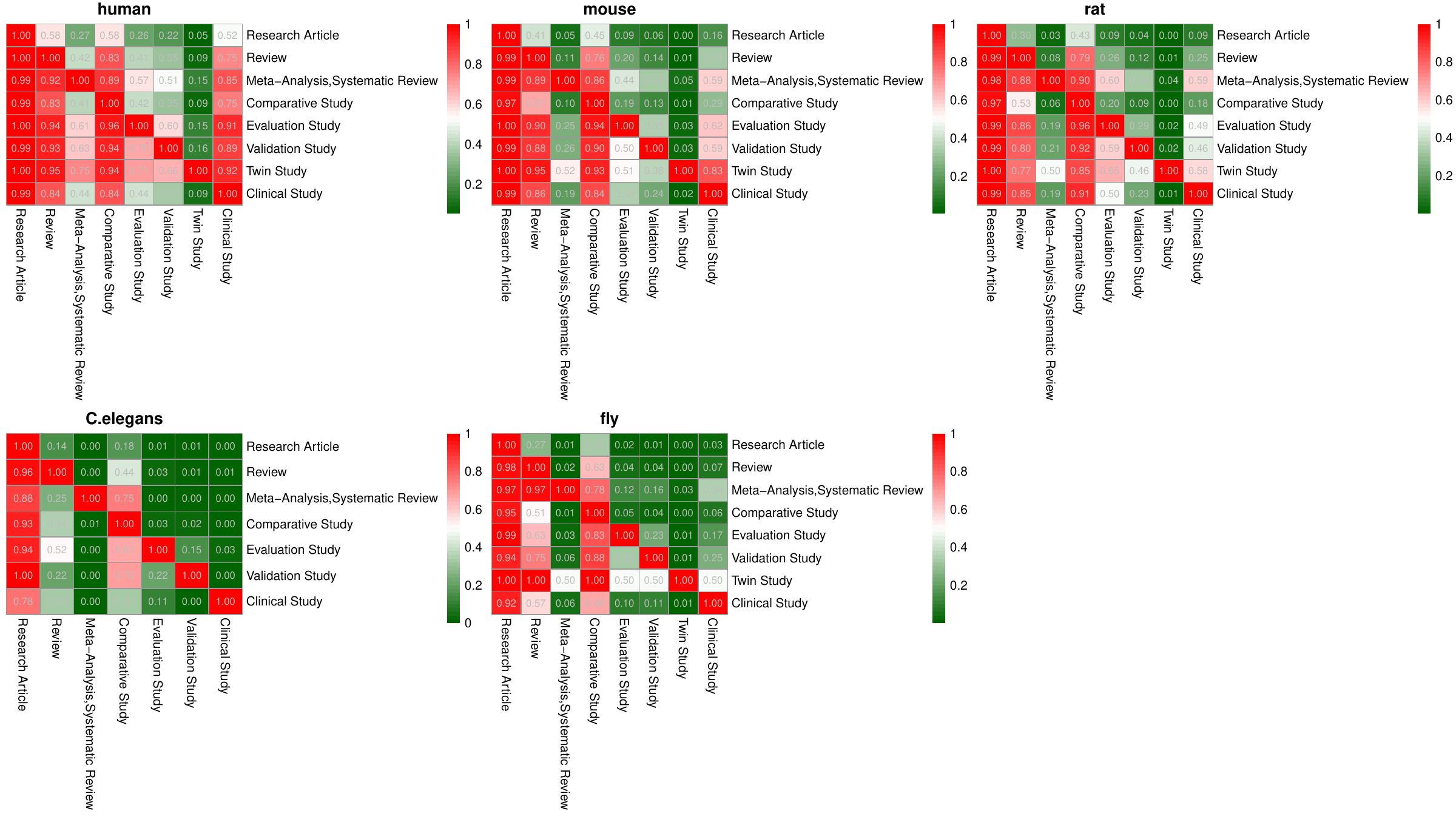}
    \caption{The heatmaps showing the overlaps of the scopes of gene studied in different publication types for Human (A), Mouse (B), Rat ( C), \textit{C.elegans}(D) and Fly (E). The elements in heatmaps represent the proportion of the intersection of genes studied by the two publication types (indexed by the row and the column) concerning the total genes studied by one publication type (indexed by the row). 
}
    \label{figure5}
\end{figure}

We took a closer look at the scopes of genes studied by different publication types and then measured the overlaps of the scopes of genes studied in different publication types as displayed in Figure \ref{figure5} and Figure \ref{figureS4}. There were differences between the scopes of genes investigated across the selected five species. For human, 3463 of over 20000 total genes were studied by research articles, comparative studies, reviews, evaluation studies, meta-analyses \& systematic reviews, compared to the numbers of overlapped genes for the rest 4 species, among which the overlap of scopes of the five publication types in rat, which is 418 in number, is highest among these four species (Figure \ref{figureS4}). 

For the five selected species in this study, research articles covered the vast majority of genes of the species studied for aging-related research (Figure \ref{figure5}, Figure \ref{figureS4}). reviews and comparative studies focused on a large subset of genes investigated in research articles, which are in 60$\sim$40\% in the number of genes studied by research articles for human and mouse while 30$\sim$10\% for \textit{C.elegans} and Fly. Comparatively, meta-analyses \& systematic reviews, evaluation studies, and validation studies showed a more concentrated focus on shrinking scopes of genes (Figure \ref{figure5}, Figure \ref{figureS4}). 

A considerable portion of genes studied by meta-analyses \& systematic reviews, evaluation studies, validation studies, and clinical studies were contained in the attention scope of reviews and comparative studies, for the four species except \textit{C.elegans} (Figure \ref{figure5}, Figure \ref{figureS4}). 

\begin{comment}
    Interestingly, Comparative Study had a relatively low percentage of genes that are also studied by Review, which is lower than the percentages of other publication types in Review, except for Research Article (Figure \ref{figure5}).
\end{comment} 

We also found that the research effort of clinical studies was concentrated on a large proportion of genes investigated by research articles for human, which was 52\% in proportion, compared to the proportions up to 16\% for the remaining four species (Figure \ref{figure5}). Only a small fraction of genes investigated by research articles and other publication types were included in twin studies for human, mouse, rat, and fly (Figure \ref{figure5}).  However, a relatively larger proportion of genes in twin studies were covered by meta-analyses \& systematic reviews (Figure \ref{figure5}). 

We also performed a \textit{post hoc} analysis to figure out whether the divergence of attention paid to aging-related genetic research across publication types was contributed by the biological natures of these genes. We employed the RE model (Section \ref{section3.1}) to aging-related publication fractions on 3463 human genes within the overlap of research articles, reviews, comparative studies, meta-analyses \& systematic reviews and evaluation studies, compared to the results without filtering genes (Figure \ref{figureS5}). The heterogeneity patterns for publication types remained to a large extent after selecting the intersected genes. It is then reasonable to deduce that the publication type-associated protocols not only gave rise to differing genetic scope but mostly introduced divergences of attention to aging-related research.

We furthermore took a closer look at the genes preferred to be studied by showing the ten most frequently studied human genes by various publication types according to the accumulated publication numbers (Table \ref{table2}, Table \ref{tableS3}). The publication of the most frequently studied 10 genes will account for a significant portion (from 20\% to 50\% varying from publication types) of the total genetic studies (Figure \ref{figure6}), as is hypothesized in Hypothesis \ref{hypothesis3} and further discussed in Section \ref{imbalance}.

\begin{table}[!htbp]
\centering
\caption{The top 10 human genes which were intensively studied in aging-related research according to the aging-related publication counts on these genes across 8 major publication types}
\begin{tabular}{rll}
  \hline
 Rank & Research Article & Review \\ 
  \hline
1 & tumor protein p53 & insulin \\ 
  2 & insulin & insulin-like growth factor 1 \\ 
  3 & cyclin-dependent kinase inhibitor 2A & tumor protein p53 \\ 
  4 & galactosidase beta 1 & growth hormone 1 \\ 
  5 & apolipoprotein E & amyloid beta precursor protein \\ 
  6 & interleukin 6 & sirtuin 1 \\ 
  7 & H3 histone pseudogene 16 & mechanistic target of rapamycin kinase \\ 
  8 & insulin-like growth factor 1 & cyclin-dependent kinase inhibitor 2A \\ 
  9 & cyclin-dependent kinase inhibitor 1A & microtubule-associated protein tau \\ 
  10 & tumor necrosis factor & apolipoprotein E \\ 
   \hline
 Rank & Comparative Study & Meta-Analysis,Systematic Review \\ 
  \hline
1 & insulin & apolipoprotein E \\ 
  2 & insulin-like growth factor 1 & forkhead box O3 \\ 
  3 & interleukin 6 & interleukin 6 \\ 
  4 & tumor necrosis factor & insulin \\ 
  5 & apolipoprotein E & insulin-like growth factor 1 \\ 
  6 & growth hormone 1 & parathyroid hormone \\ 
  7 & CD4 molecule & CD4 molecule \\ 
  8 & amyloid beta precursor protein & C-reactive protein \\ 
  9 & tumor protein p53 & paraoxonase 1 \\ 
  10 & CD8a molecule & CD8a molecule \\ 
   \hline 
   Rank & Evaluation Study & Validation Study \\ 
  \hline
1 & hemoglobin subunit alpha 1 & anti-Mullerian hormone \\ 
  2 & insulin & apolipoprotein E \\ 
  3 & cut like homeobox 1 & insulin-like growth factor 1 \\ 
  4 & selenium binding protein 1 & TNF superfamily member 10 \\ 
  5 & ABL proto-oncogene 1 & poly(ADP-ribose) polymerase 1 \\ 
  6 & anti-Mullerian hormone & cytochrome c oxidase subunit 8A \\ 
  7 & caspase 3 & C-reactive protein \\ 
  8 & CD34 molecule & dedicator of cytokinesis 3 \\ 
  9 & D-box binding PAR bZIP TF & FAT atypical cadherin 1 \\ 
  10 & MYC proto-oncogene & microtubule-associated protein tau \\ 
   \hline
Rank & Twin Study & Clinical Study \\ 
  \hline
1 & apolipoprotein E & insulin \\ 
  2 & insulin & insulin-like growth factor 1 \\ 
  3 & interleukin 6 & growth hormone 1 \\ 
  4 & apolipoprotein A1 & interleukin 6 \\ 
  5 & apolipoprotein B & C-reactive protein \\ 
  6 & C-reactive protein & tumor necrosis factor \\ 
  7 & angiotensin I converting enzyme & CD59 molecule (CD59 blood group) \\ 
  8 & tumor necrosis factor & growth hormone releasing hormone \\ 
  9 & E1A binding protein p300 & apolipoprotein E \\ 
  10 & insulin like growth factor 2 & parathyroid hormone \\ 
  \hline
  
\end{tabular}
\label{table2}

\end{table}

Despite the resemblance between research articles and reviews, research articles put extra concentration on genes that are more studied in the fundamental context of biology on molecular mechanisms of cellular aging (Table \ref{table2}), for instance, galactosidase beta 1 (BETA1), H3 histone pseudogene 16 (H3H16P), and cyclin-dependent kinase inhibitor 1A (CDK-inhibitor 1A). Evaluation studies and validation studies show distinctive studying preferences with little overlap with other publications (Table \ref{table2}). However, this substantial shift in evaluation studies became weaker using citation counts (Table \ref{tableS3}). One plausible explanation is that the total publication counts of these two publication types are very limited while citation counts are fair enough to avoid statistical instability (Section \ref{appendix_data_retrieval} and \ref{appendix_citation}). As another major publication type, clinical studies concentrated on genes that have higher clinical significance and are easy to detect in surgery and clinical practice, including multiple kinds of hormones, such as insulin, growth hormone 1 (GH1), growth hormone-releasing hormone (GHRH), or biomarkers like C-reactive protein (CRP). 

Although shared a large overlap with other publication types,  meta-analyses \& systematic reviews especially had frequently studied genes uncovered by the predominantly investigated genes in other publication types, such as CD4 and CD8a, which play important roles in age-related maladies and disorders, forkhead box O3 (FOXO3) and parathyroid hormone (PTH). Twin Study also had several genes not listed by others, including apolipoprotein B (ApoB), apolipoprotein A1 (ApoA1), angiotensin I converting enzyme (ACE), insulin growth factor 2 (IGF2), and E1A binding protein p300 (p300) which could be measured directly from serum.%CD59, a protein related to age-related immune system decline, was put extra emphasis on by Clinical Study.

Though differences between publication types exist, it was notable that some genes were put enough emphasis on by all publication types. For example, the growth-associated genes coding the insulin and the insulin-like growth factor (IGF) kept the most popular positions across the selected eight publication types.  Research articles and reviews showed a strong resemblance between preferred areas. The Apolipoprotein E (ApoE) which is associated with lipid metabolism and Alzheimer’s disease draws the most intensive attention by all publication types except evaluation studies. 
These most frequently studied genes are involved with multiple aging-related topics including metabolism, cellular pathways, oncology, development, and immunology. Publication types showed varying interests in human genes and the aging-related topics behind them (Table \ref{table2}). 

\subsection{The imbalance of attention on genes by each publication type}\label{imbalance}
As suggested by recent research that there exists a highly imbalanced research effort prevailing in biomedical research to individual human protein-coding genes \parencite[]{stoeger2018large,haynes2018gene,stoeger2022characteristics}, we hypothesized on the imbalance of bibliometric data on biomedical topics across types of publications (Hypothesis \ref{hypothesis3}). 
\begin{figure}[!htbp]
    \centering
    \includegraphics[width=1\columnwidth]{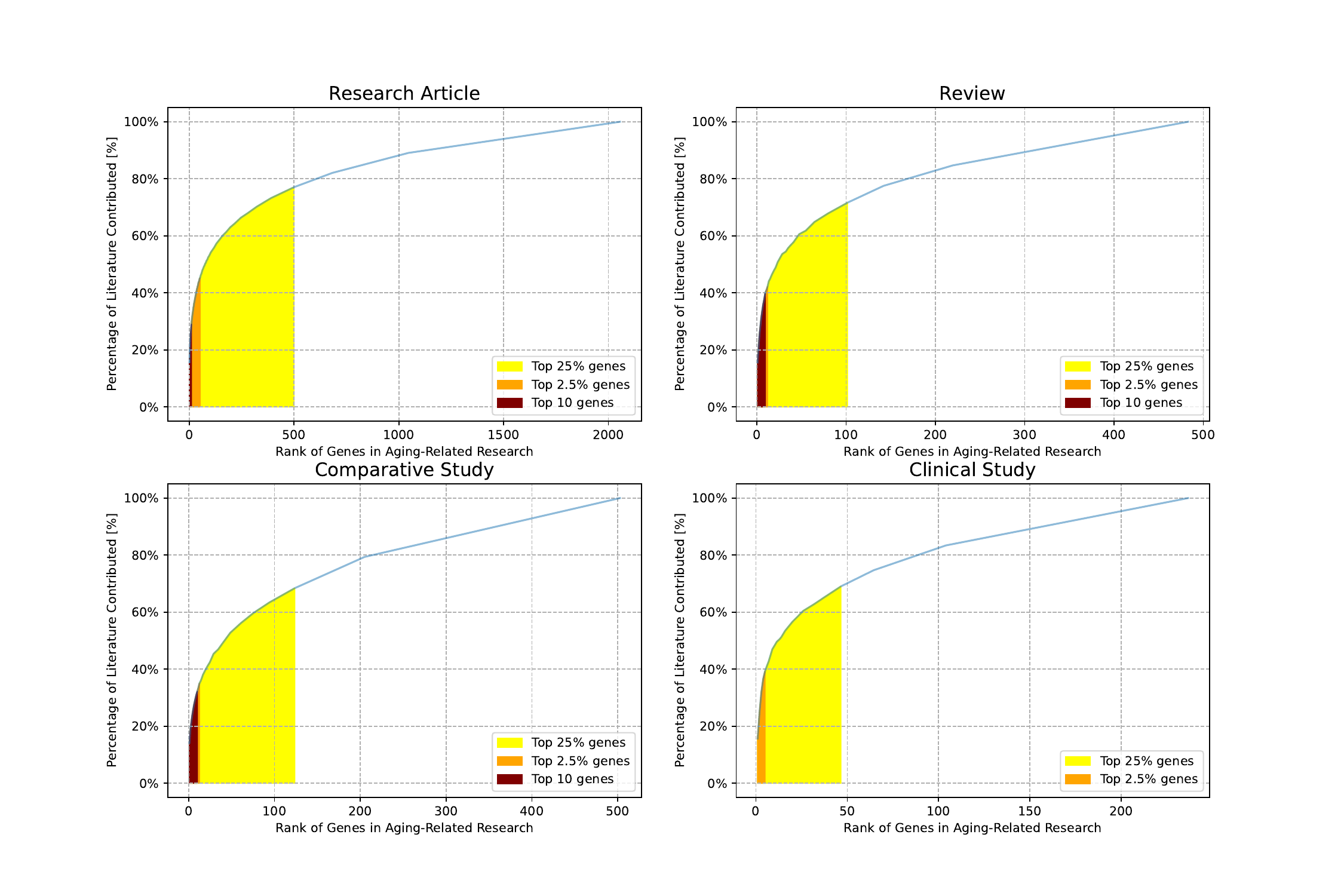}
    \caption{Cumulative share of aging-related literature on human genes tagged in the filtered publications from PubMed for the four most common publication types (research articles, reviews, comparative studies, and clinical studies). Gene rank refers to the order of human genes according to the associated cumulative publication numbers. The gene with the most publication equivalents would have ranked 1. The share of total literature contributed by top 25\%, 2.5\%, and 10 genes are shaded. Note: for the fourth sub-figure the share of top ten genes is not shaded because the number 10 exceeds the number of 2.5\% of total genes involved.
    \label{figure6}
}
\end{figure}
Partially supporting our hypothesis, the imbalance of research efforts among four major types is observed (Figure \ref{figure6}), as well as the topics and genes preferred to be unevenly investigated for each type (Figure \ref{figure4} and Table \ref{table1}). However, counter-intuitively, we cannot observe obvious differing patterns in the distribution of research attention across types from Figure \ref{figure6}, because it is reasonable to assume a larger extent of imbalance for reviews, comparative studies, and clinical studies as reviews and application studies should be developed upon the discovery by liminal research studies. It may suggest a complexity in the interchange and dissemination of knowledge between types in this specific field, which will be further developed in Section \ref{discussion}.%Supporting our Hypothesis \ref{hypothesis3}, we find that not only over 30\% publications targeted only 10 genes, but the portion of literature unevenly contributed by top genes dramatically changes across publication types (Figure \ref{figure6}). Specifically, %the 40 top ranked genes represent a little over 1% of the newly highlighted genes, but 21% of all unique

\subsection{Bibliometric variations among publication types further characterize the intricacies}
By Hypothesis \ref{hypothesis4} we may infer the divergence in the bibliometric dimensions of the publication data. In this subsection, we provide data supporting our hypothesis and argue that given bibliometric metrics among publication types, we can further characterize the roles and functions of specific types in knowledge dissemination.

We first argue that aging-related publications demonstrate certain patterns of preferences in some journals regarding or regardless of the publication types (Table 3 and Figure \ref{figureS6}). The top 10 journals ranked by aging-related publication counts for each publication type are exhibited first (Table 3), as well as the top 10 journals with the highest proportions on aging-related studies among 8 publication types with Bnferroni-corrected p-values (Table \ref{tableS4}). 

\begin{table}[!htbp]
    \label{table4}
    \centering
    \scalebox{0.8}{\begin{tabular}{rllll}
\hline
 Rank &                                   Research Article &  Count &                                             Review &  Count\\
\hline
1 &                                           PloS one &                           379 &                            Ageing research reviews &                  90 \\
2 &                              Neurobiology of aging &                           286 &        IJMS &                  84 \\
3 &                                         Aging cell &                           278 &                           Experimental gerontology &                  80 \\
4 &                                              Aging &                           275 &               Mechanisms of ageing and development &                  72 \\
5 &               Mechanisms of ageing and development &                           205 &         Annals of the New York Academy of Sciences &                  48 \\
6 &  JCEM &                           205 &                                              Cells &                  42 \\
7 &                                 Scientific reports &                           194 &                                              Aging &                  33 \\
8 &                           Experimental gerontology &                           188 &      Advances in experimental medicine and biology &                  29 \\
9 &  PNAS &                           157 &  Nihon rinsho. Japanese J. of clinical med. &                  29 \\
10 &  BBRC &                           155 &                                        Gerontology &                  26 \\
\hline

Rank &                                  Comparative Study &  Count &                  Meta-Analysis \& Systematic Review &  Count \\
\hline
0 &  JCEM &                             70 &                            Ageing research reviews &                                             10 \\
1 &                              Neurobiology of aging &                             50 &                           Experimental gerontology &                                              7 \\
2 &  The journals of gerontology. Series A &                             46 &                                           PloS one &                                              6 \\
3 &      J. of immunology (Baltimore) &                             43 &                              Rejuvenation research &                                              6 \\
4 &               Mechanisms of ageing and development &                             42 &  The journals of gerontology. Series A &                                              4 \\
5 &                           Experimental gerontology &                             33 &                                         Aging cell &                                              4 \\
6 &                       Age (Dordrecht, Netherlands) &                             30 &         Zeitschrift fur Gerontologie und Geriatrie &                                              3 \\
7 &                                              Blood &                             21 &               Mechanisms of ageing and development &                                              3 \\
8 &                                     Brain research &                             21 &                          Human reproduction update &                                              3 \\
9 &                                 Clinical chemistry &                             20 &                              Nature communications &                                              3 \\
\hline

Rank &                                   Evaluation Study &  Count &                                   Validation Study &  Count \\
\hline
1 &           Aging clinical and experimental research &                             2 &            The J. of nutritional biochemistry &                             3 \\
2 &         The American J. of surgical pathology &                             2 &                      BioMed research international &                             2 \\
3 &  Quality of life research  &                             2 &                              Rejuvenation research &                             2 \\
4 &                                 Acta diabetologica &                             1 &                                           PloS one &                             2 \\
5 &                   Pigment cell \& melanoma research &                             1 &  J. of applied physiology  &                             2 \\
6 &                                      Klinika oczna &                             1 &                               Molecular immunology &                             2 \\
7 &                                          Maturitas &                             1 &               Acta neuropathologica communications &                             1 \\
8 &                       Medicina (Kaunas, Lithuania) &                             1 &                       Medicina (Kaunas, Lithuania) &                             1 \\
9 &                                 NMR in biomedicine &                             1 &  JCEM &                             1 \\
10 &                               Neuroscience letters &                             1 &                  The American J. of pathology &                             1 \\
\hline

Rank &                                         Twin Study &  Count  &                                     Clinical Study &  Count\\
\hline
1 &  JECM &                       7 &  JECM &                         130 \\
2 &                                              Aging &                       2 &              Metabolism: clinical and experimental &                          28 \\
3 &                                       Diabetologia &                       2 &                  European J. of endocrinology &                          28 \\
4 &  The journals of gerontology. Series A &                       2 &                                           PloS one &                          25 \\
5 &               The J. of experimental medicine &                       2 &                              Neurobiology of aging &                          23 \\
6 &                                           PloS one &                       2 &  J. of applied physiology &                          18 \\
7 &                                   Neuropsychologia &                       2 &               Mechanisms of ageing and development &                          17 \\
8 &                                             Nature &                       2 &                                   Hormone research &                          17 \\
9 &                                          Neurology &                       2 &         The American J. of clinical nutrition &                          16 \\
10 &  American J. of medical genetics. Part B &                       2 &  The journals of gerontology. Series A &                          15 \\
\hline
\end{tabular}
}
\caption{The top 10 journals with the highest publication counts across 8 different publication types. BBRC:        Biochemical and Biophysical Research Communications; PNAS: Proceedings of the National Academy of Sciences of the United States of America; IJMS:  International J. of Molecular Sciences; JCEM: The J. of Clinical Endocrinology and Metabolism}
\end{table}

From the perspective of subject specialty and audience scope of journals, different article types match journals of various natures. Research articles tend to be published in high-impact multidisciplinary journals, such as PLoS One and PNAS, and are also seen in aging-related specialized journals, such as Experimental Gerontology.  In contrast, reviews prefer more specialized journals (and some journals specializing in reviews), such as Mechanisms of Ageing and Development, and are also present in high-impact interdisciplinary journals, such as IJMS. Comparative studies mainly target specialized medical journals, with some present in general journals. Except for those 6 publications on PloS one, meta-analyses and systematic reviews are only published in medical journals specializing in particular domains, such as Rejuvenation Research and Experimental Gerontology. Evaluation studies and validation studies are limited to specialized and topical medical publications. Twin studies and clinical studies have a relatively wider distribution and are also present in general science journals.

%Specialized gerontology journals accommodate diverse article types. 
In addition, specialized journals concentrate on specific research issues, while general journals cover a wider range of topics. Research articles cover a wide range of topics, from fundamental landscapes such as cell biology and genetics to clinical research such as pharmacology and pathology. Reviews concentrate on gerontology themes, such as mechanisms of aging and anti-aging interventions, with some clinical reviews included. In contrast, comparative studies focus on disease models and mechanisms. Evaluation studies are concerned with quality assessment and psychometric research. Validation studies focus on technique and methodology validation. Twin studies are primarily in the field of genetics. Clinical studies mainly involve clinical trials and treatments.
%Meta-analyses and systematic reviews emphasize quantitative research synthesis.

\begin{figure}[!htbp]
    \centering
    \includegraphics[width=1\columnwidth]{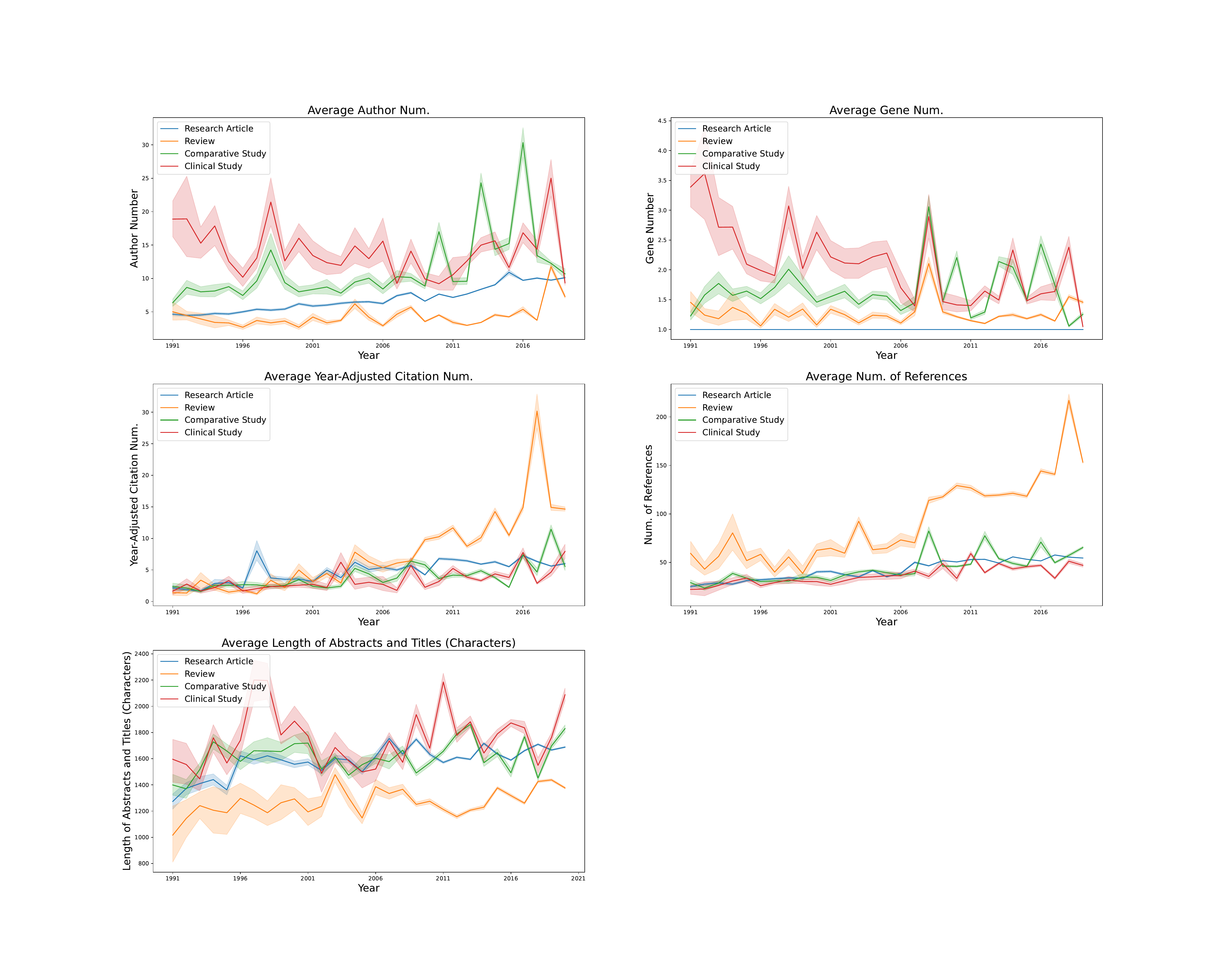}
    \caption{The average values of five frequently used bibliometric measures for aging-related research articles, reviews, comparative studies, and clinical studies on human genes entered from 1991 to 2020. The borders of shaded areas indicated the 95\% confidence intervals of estimation by bootstrap.
}
\label{figure7}
\end{figure}

\begin{figure}[!htbp]
    \centering
    \includegraphics[width=1\columnwidth]{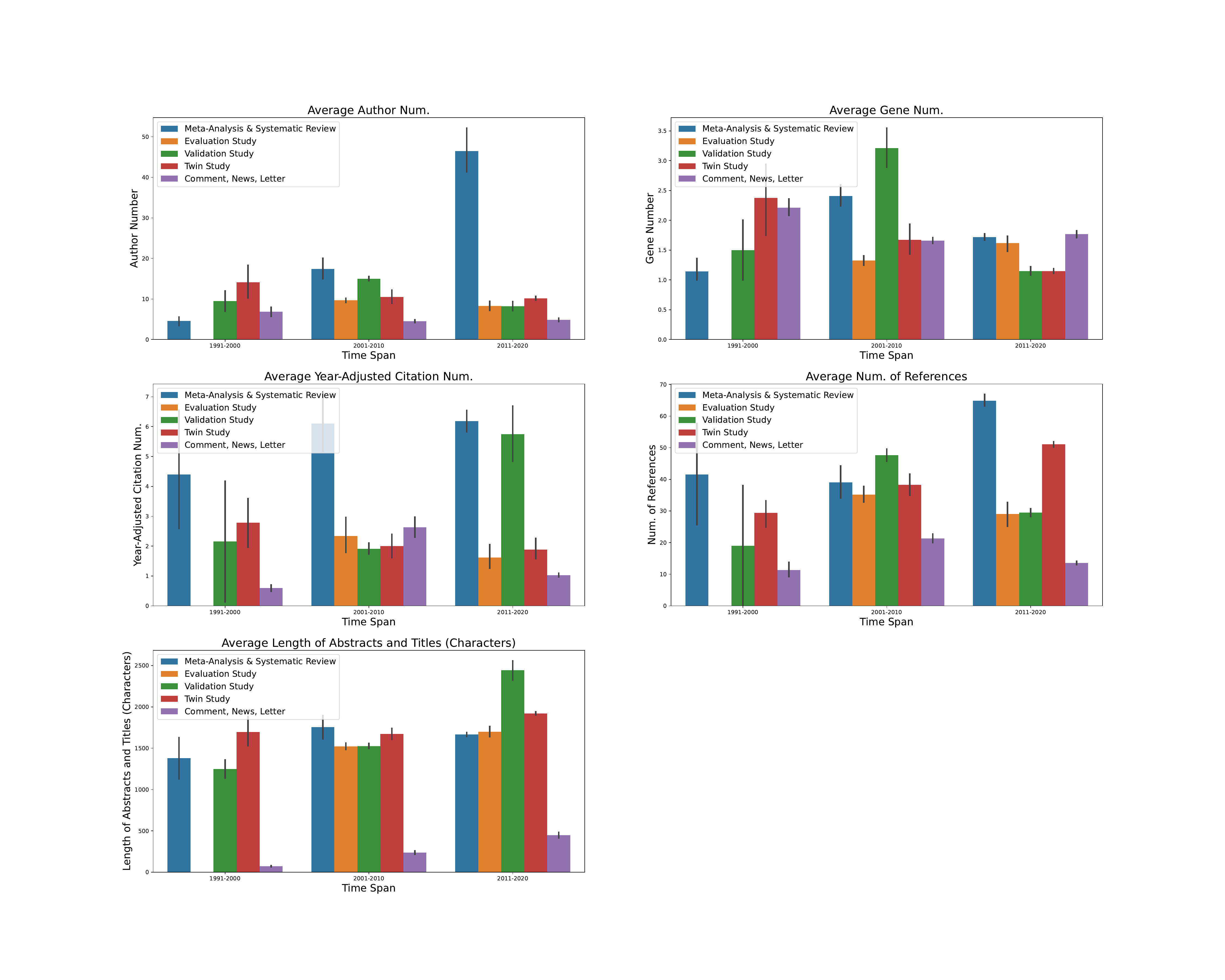}
    \caption{The average values of four frequently used bibliometric measures for aging-related meta-analyses \& systematic reviews, evaluation studies, validation studies, twin studies, and comments \& news \& letters on human genes entered from 1991 to 2020.
}
\label{figure8}
\end{figure}

In addition, the four major publication types (research articles, reviews, comparative studies, and clinical studies) manifested different patterns and trends in commonly used scientometric measures (Figure \ref{figure7}), as well as for the other five minor types as shown in Figure \ref{figure8}. 

Except for clinical studies and reviews, the number of authors for the other two publication types showed an overall upward trend, with the author number for research articles steadily increasing (Figure \ref{figure7}). Over the three decades from 1991-2020, the average number of genes studied per article for clinical studies declined, while the other three publication types remained relatively stable (Figure \ref{figure7}). The length of titles and abstracts showed slightly increasing trends for four publication types. The average annual citation increment and average reference counts also increased for all types (Figure \ref{figure7}). Among these four major types, reviews had the fastest growth in citation count, while the other three types grew at comparable levels (Figure \ref{figure7}). Unlike hotspot research areas, reviews had the fewest authors, followed by research articles, while comparative studies and clinical studies had the most authors (Figure \ref{figure7}). Reviews contained the greatest number of references, with the other three types growing at similar speeds. For reviews, both the mean yearly citation increment and average citation count underwent rapid growth around 2008 (Figure \ref{figure7}).

The average length of abstracts and titles showed a steady increase for all five minor publication types (Figure \ref{figure8}). Meta-analyses \& systematic reviews, compared to the other four minor article types, showed a significant rise in the average number of authors over three decades (Figure \ref{figure8}). Also, meta-analyses \& systematic reviews received more average annual citation increments compared to the other four minor types, with the average annual citation counts increase for validation studies rising after 2010 (Figure \ref{figure8}). The average reference counts also increased over three decades for both meta-analyses \& systematic reviews and twin studies (Figure \ref{figure8}).

\begin{figure}[!htbp]
    \centering
    \includegraphics[width=1\columnwidth]{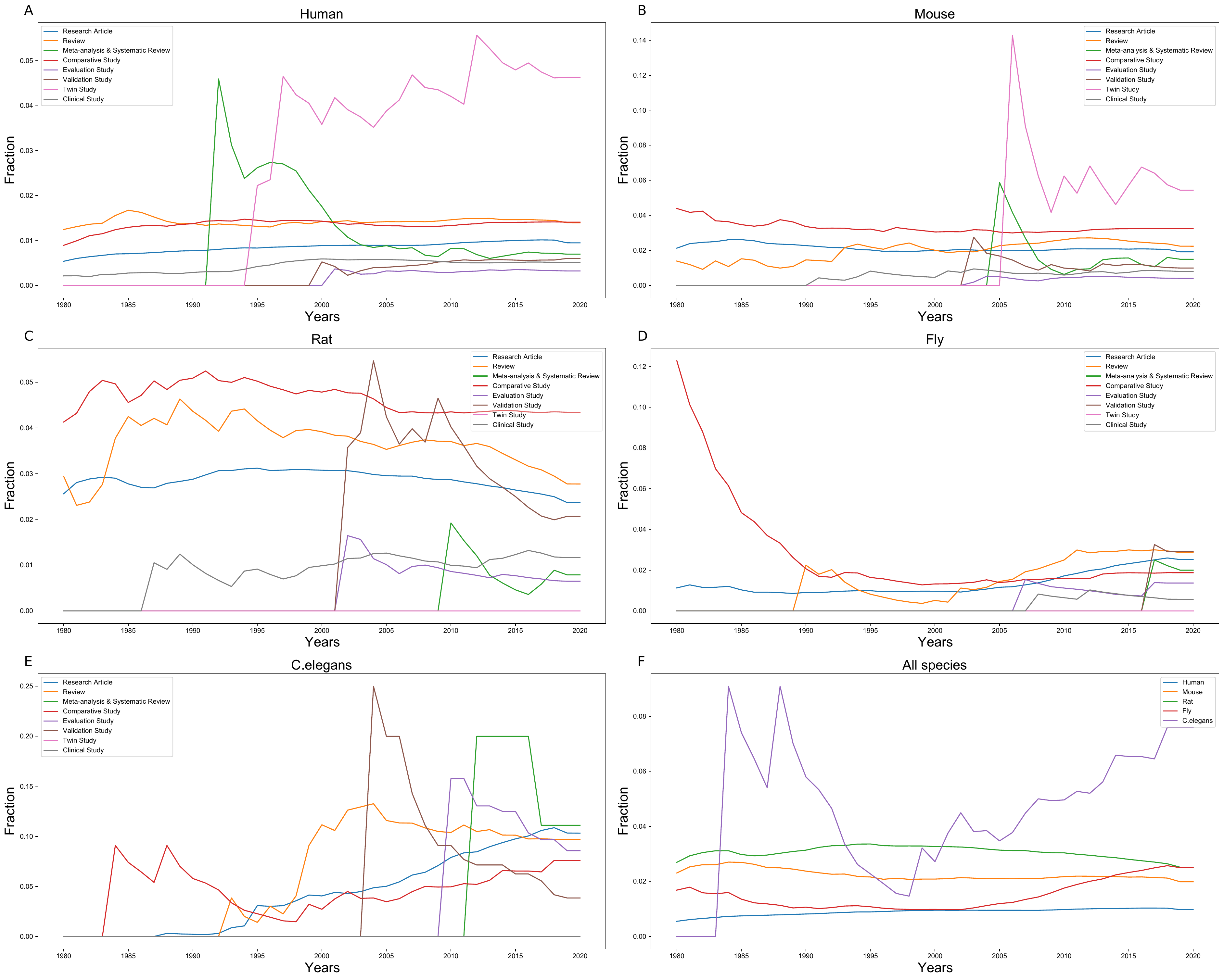}
    \caption{The temporal trends of the publication fractions of aging-related research by 8 publication types (research articles, reviews, comparative studies, meta-Analyses \& systematic reviews, evaluation studies, validation studies, twin studies, and clinical studies) from 1980 to 2020. The publication fractions were calculated by retrieving the recorded publications from the PubMed data sets. }
    \label{figure3}
\end{figure}

We also performed a sensitivity analysis to inspect how the aging-related research fractions changed over the past four decades. As shown in Figure \ref{figure3}, we found there are substantial differences in the temporal trends of attention to aging-related research for various publication types. Research articles, comparative studies, reviews, and clinical studies remained stable on the fractions of aging-related publications over 1980-2020 across five selected species (the seemingly wiggling trends of Rat, Fly, and \textit{C.elegans} is due to the limited sample sizes). Meta-analyses \& systematic reviews, twin studies, evaluation studies, and validation studies on aging were initiated during the 1990s and the aging-related fractions of these four publication types evolved to stable status thereafter. 
The growth rate of the number of papers in different publication types can somewhat reveal the current focus of development in a field. Compared to the around three times growth of the annual publication number of research articles, the rapid growth in the annual publication number of meta-analyses and systematic and the fluctuations of annual publication counts for the rest six types in aging-related literature is worth concern (Figure \ref{figure3}, Table \ref{tableS2}, Figure \ref{figureS3}). Both meta-analyses and systematic reviews are powerful in assessing the correctness and effectiveness of theories and treatments by synthesizing studies and identifying trends unapparent in individual studies \parencite[]{shenkin2017systematic}, and thus contain higher levels of evidence with more stringent statistical thresholds and fewer possibilities of being false positive. The increasing trend of meta-analyses and systematic reviews may indicate a higher acceptance of research with higher levels of evidence for the academic communities. It was also notable that the annual publication numbers of comparative studies and clinical studies expanded less rapidly than research articles and reviews for the human and the mouse (Figure \ref{figureS2}).%defififufu

\section{Discussion and Conclusion}\label{discussion}
This paper investigated and compared attention in the aging-related domain on genes across 12 publication types in 5 species. Using the fraction of aging-related publications and citation counts for each gene, we quantified the attention to aging-related research relative to nonaging-related research by applying a RE model. Moreover, we investigated the scopes, overlaps, and inequality of research efforts of genes studied by various publication types for aging-related research. This study aimed to indicate diverse genetic research interests impacted by different publication protocols in the context of aging-related research. 
\subsection{Publication types serve for naturally different purposes and investigate intertwining scopes}
The exhibited data indicates naturally different aims for each type from several perspectives. First, each type pays uneven attention to one particular field. It was commonly acknowledged that the variations of attention to a particular field vary across different publication types. Consistent with previous investigations \parencite[]{shenkin2017systematic}, we demonstrated that the meta-analyses and systematic reviews focused more on aging-related research on genes (Figure \ref{figure1}). Additionally, the comparative studies showed the richest attention to aging-related genetic research (Figure \ref{figure1}) among seven publication types (Figure \ref{figure2}). The characteristic of comparative studies covering multiple species and lifespan intervals in aging-related fields \parencite[]{austad2009comparative} might have contributed to the preference of this publication type by researchers. Interestingly, attention to aging-related genetic research was the lowest in evaluation studies compared to other publication types (Figure \ref{figure1}, Figure \ref{figure2}). 

We further examined the potential contributors to the diverging aging-related attention by comparing the average aging-related research fractions across publication types and species before and after performing the RE model (Section \ref{section3.1}) on the overlap of genetic scopes among five publication types (Figure \ref{figureS5}). It was plausible to infer from the confirmatory results in Figure \ref{figureS5} that the various publication type-associated factors (such as the protocols conducting research), rather than the different biological nature of genes concentrated on by different publication types, resulted in the inconsistency of aging-related attention. 

Secondly, the differences and similarities between the genetic scopes in various publication types help us understand how the publication type-associated factors influence the preferences of genes. Figure \ref{figure4} suggests methodological preferences and object-level variation among four types of publications, from genotypic technologies to clinical investigations. Table \ref{table2} provides further evidence that these types concentrate on different branches of biomedical sciences ranging from the context of cellular fundamentals to clinical hormones studies. Furthermore, the frequencies of publications in Table 3 signify various scopes and levels of readerships and specialization of journals among types of publications.

Different aging-related genetic topics received varying levels of attention across publication types. From Table \ref{table2} and Table \ref{tableS3}, we can observe that the eight displayed publication types focused on some common topics, including the regulation (insulin and IGF1) and immunology (interleukin 6) of aging processes \parencite[]{guarente2000genetic,anisimov2013key,michaud2013proinflammatory}. The ApoE which involved neurodegenerative and cardiovascular diseases \parencite[]{yin2018apoe} received the richest level of concern among all publication types. The research articles put more emphasis on CDK-inhibitor 1A and 2A, p53, and BETA1 which are associated with the molecular mechanisms and identification of cellular senescence \parencite[]{casella2019transcriptome,hernandez2017unmasking}. The CD4 and CD8a involved in the immunological responsiveness to carcinogenic cells and exogenous \parencite[]{haynes1997age} components were intensively focused on by comparative studies, meta-analyses, and systematic reviews. Consistent with the methodologies and objectives of evaluation studies and validation studies which aim to evaluate and benchmark the effectiveness of procedures, these two types of studies focus more on the aging biomarkers whose levels reflect multiple aspects of aging processes, such as caspase 3 \parencite[]{abu2001biomarkers}, CD34 \parencite[]{franceschi1999biomarkers} and CRP \parencite[]{niculescu2000identifying}. Numerous studies of these two types were also involved with genetic markers of several classes of aging. For example, the changes in levels of the hemoglobin subunit alpha 1, anti-Mullerian hormone, and selenium binding protein 1 are associated with various age-related diseases \parencite[]{chao2012anti,wu2012selenium}. The twin studies concentrated more on molecules that are accessible by clinical techniques and lead to or suggest severe diseases. More concretely, ApoB, ApoA1, p300, and ACE are associated with age-related cardiovascular diseases, cancers, and cognitive decline \parencite[]{caramelli1999increased,shilpasree2013study}. Remarkably, the genes received unbalanced citation frequencies depending on the topics. Comparing Table \ref{table2} and Table \ref{tableS3}, we can observe that clearly insulin, IGF1, and interleukin 6 impacts the upcoming studies disproportionately compared to the publication data. The imbalance is conspicuous for evaluation studies. This evidence might suggest that the historical reasons such as the landmarking impacts inspired and promoted by the research on insulin, IGF1, and interleukin 6 exerted disproportionate influence compared to its actual importance on these genes in the evaluation studies.

Third, the different topology between scopes of genes investigated across species may suggest dissimilar research patterns (Figure \ref{figure5}). We observed that half aging-related genes studied in the research articles were included in clinical studies, while the percentage for the rest four species was under 20\% (Figure \ref{figure5}). This is consistent with the objectives of clinical studies that aim to improve the treatments and interventions for human healthcare. We also found an interesting result: the divergent attention genetic scopes of comparative studies and reviews across five species, suggested by the relatively lower percentages of genes studied in comparative studies also investigated in reviews (Figure \ref{figure5}).  This may suggest barriers and less smooth knowledge diffusion between reviews and comparative studies.

\subsection{Complex steering forces behind knowledge production and dissemination}
\begin{figure}[!htbp]
    \centering
    \includegraphics[width=1\columnwidth]{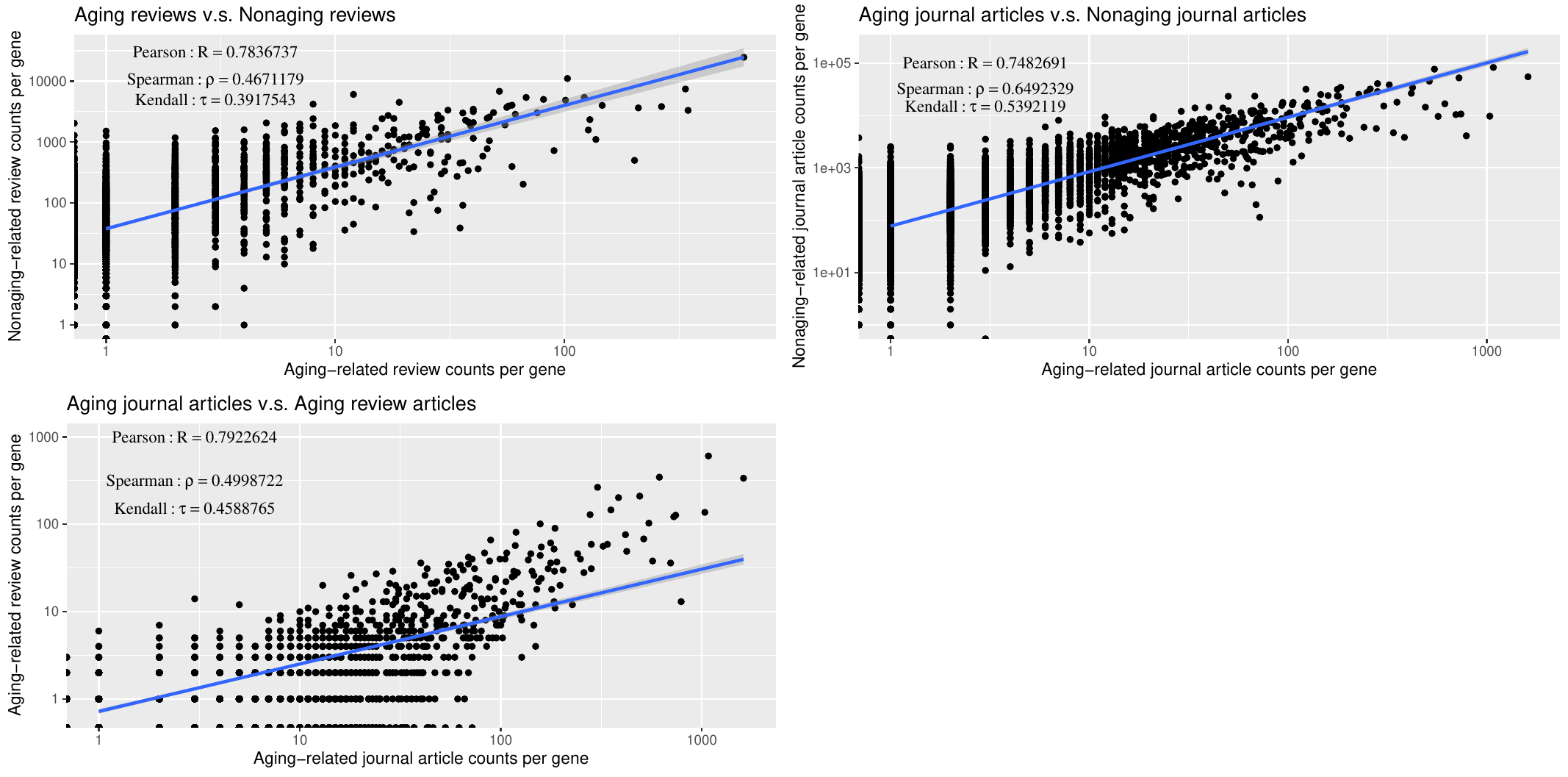}
    \caption{Scatter plots of the publication counts of aging-related review counts \textit{v.s.} nonaging-related review counts, aging-related journal article counts \textit{v.s.} nonaging-related journal article counts, and aging-related journal article counts \textit{v.s.} aging-related review counts. Pearson's, Spearman's, and Kendall's correlations are labeled.}
\label{figure9}
\end{figure}

The data from Figure \ref{figure6} in Section \ref{imbalance} suggests the hidden complexity of knowledge dissemination and production processes among publication types, as embodied by the similar levels of imbalance of attention. Here we closely examine the assumption that reviews and application studies (such as comparative studies and clinical studies) are built upon original studies (such as research articles) and thus are more imbalanced into targeted topics with special significance. From Figure \ref{figure9} we can see that the counts of aging-related reviews on each gene are highly correlated with both aging-related journal articles (including all articles via journals) and nonaging-related reviews. Therefore, we may argue that aging-related reviews are to some extent developed by referencing the boarder scope of aging-related studies. Nevertheless, the correlation between aging-related and nonaging-related journal articles and reviews in Figure \ref{figure6}, together with the similar levels of imbalance in Figure \ref{figure6}, implies more complicated contributors to knowledge production and dissemination in biomedical sciences. Knowledge production in biomedical science, which is at the core of scientific communication, may not only be dependent on natural and scientific significance.
\subsection{In sum: a finer characterization of knowledge production and dissemination in a subdomain}

In this article, we performed a comprehensive quantitative analysis of the publication data of aging-related subdomains regarding different publication types. Our data signifies different patterns of research attention and overlapping scopes of biomedical topics. Furthermore, we reproduced the scientometric data and investigated the imbalance regime of research efforts across publication types. By applying various quantitative methods and examining data on several dimensions, our analysis is embodied to convey a finer characterization of the roles and functions of each publication type in the biomedical knowledge production systems. 

This study has several limitations. To begin with, although we adopted the advocated MeSH-based search strategy \parencite[]{leydesdorff2012bibliometric,leydesdorff2013citation}, not all publications recorded were tagged with appropriate labels summarizing the information of publication types by MeSH. In addition, some of the current publications were not indexed by PubMed and thus excluded from this study. The vague boundaries among the publication types, such as comparative studies and clinical studies (including the original MeSH categories), are worth noting. Also, there is the possibility that articles were manually tagged as the adjacent publication types, for example, one systematic review may be tagged as “Review”. In this study, we quantify the attention to aging-related research using the RE model, which requires normality (or close normality) because of its statistical nature. However, we observed the long-tailed distribution of the publication and citation fractions on genes, which was inconsistent with the model assumptions. 

We depicted the attention patterns in biomedical research of diverse publication types using publication information from PubMed. The elevation of large-scale bibliometric data provides a basis for a thorough and objective understanding of attention in one field. The enhanced understanding of the attention pattern differences between publication types should be able to help promote scientific communication across communities holding varying research ideologies and methodologies. Moreover, the presenting study could also be extended by including more dimensions of attention in scientific research. This study could be regarded as a start for future monitoring of any biomedical domain from the perspective of publication types.

Practically speaking, learning how publication type-associated factors influence knowledge production and diffusion is helpful for policy-makers to evaluate the scientific significance of related research. The data-driven approach guarantees a more evidence-based view for policymakers, which in turn also leads to mutual understanding between scientists and policymakers, and a more debatable environment \parencite[]{choi2005can}. Our result and discussion also provide a unique perspective for researchers in need of a deeper understanding of not only the scientific knowledge of one field but also the matters of science conducting.

\appendix
\section{Appendices}
%\begin{appendices}

\subsection{Detailed notes on bibliometric data retrieval}\label{appendix_data_retrieval}
We restricted our analysis to publications in the biomedical field. Publication information including PubMed ID, publication type, published date, and MeSH \parencite[]{dhammi2014medical} terms tagged was extracted from the PubMed database. PubMed data was downloaded from \url{https://www.nlm.nih.gov/databases/download/pubmed_medline.html}. Results of Figure \ref{figure1}, \ref{figure5}, \ref{figure5}, Table \ref{table2} and \ref{tableS2} used PubMed data downloaded on October 21, 2020. The rest of the results used data downloaded on January 14, 2023. We excluded publications published later than 2020 due to insufficient recording and editing by PubMed. Citation relationship between publications used data downloaded from iCite website \url{https://nih.figshare.com/collections/iCite_Database_Snapshots_NIH_Open_Citation_Collection_/4586573/32} \parencite[]{hutchins2019nih}. Gene information including NCBI ID, symbol, description, and taxonomy ID was obtained from \url{https://ftp.ncbi.nlm.nih.gov/gene}. Data of NCBI gene supporting Figure \ref{figure1}, \ref{figure5}, \ref{figure5}, Table \ref{table2} and \ref{tableS2} was downloaded on December 21. NCBI gene data supporting the rest of results was downloaded on August 16, 2022.

We used data constructed by PubTator to filter the genes mentioned in titles and abstracts \parencite[]{wei2019pubtator}. The data was downloaded from \url{https://ftp.ncbi.nlm.nih.gov/pub/lu/PubTatorCentral/}. We specifically used gene2pubtatorcentral.gz. PubTator data supporting Figure \ref{figure1}, \ref{figure5}, Table \ref{table2} and \ref{tableS2} was downloaded on October 22, 2020. The rest of the results used PubTator data downloaded on July 12, 2022.

The data downloaded was parsed by Gustav, which is a framework extracting and organizing bibliometric data \parencite[]{stoeger2022aging}. Gustav converted raw data to Python dataframes. The gene publication pairs of PubTator were filtered by excluding the data entries containing ";" in gene columns to avoid ambiguity. We used the following functions in gustav to construct Python dataframes for our downstream analyses: ncbi.gene\_info, ncbi.gene2pubmed, ncbi.pubmed, ncbi.pubtator\_medline, nih.icite, nlm.mesh.

In order to narrow our focus on scientific publications in the biomedical sciences, we chose the following publication types for investigations: “Review”, ”Introductory Journal Article”, ”Comparative Study”, ”Historical Article”, ”Evaluation Study”, “Validation Study”, “Twin Study”, “Classical Article”, "Journal Article", "Corrected and Republished Article", "Pragmatic Clinical Trial", "Observational Study, Veterinary", "Clinical Conference", "Clinical Study", "Clinical Trial, Veterinary", "Clinical Trial Protocol", "Clinical Trial, Phase I", "Clinical Trial, Phase II", "Clinical Trial, Phase III", "Clinical Trial", "Randomized Controlled Trial", "Case Reports", "Controlled Clinical Trial", "Observational Study", "Meta-Analysis", "Systematic Review", "Comment", "News", and "Letters". 

We observed one issue that the publication types, which contain unusual clinical study designs, showed a small number of total publication counts on aging-related genetic research. We, therefore, amalgamated these clinical publication types as a new category named "Clinical Study" as we defined in Table \ref{table1}, under the assumption that the differences in bibliometric characteristics within clinical publications are trivial. We applied similar procedures to "Meta-Analysis", "Systematic Review", "Comment", "News", and "Letters" (Table \ref{table1}). 

We have introduced a new term, "Research Article," which encompasses ordinary biomedical research publications that may not have undergone certain analyses, trials, or experiments. PubMed does not classify these publications as particular biomedical publication types, such as comparative studies or meta-analyses, which may have distinct research evaluation standards. Note that the corrected and republished articles not meeting certain classification criteria were also recognized as "Research Article." Retracted publications were not covered in the current analysis.

% and 4 classes (Table \ref{table1}) of publications which were merged from some remaining individual types because of similar research objectives and methodologies. 

We chose 5 representative species intensively studied both inside and outside aging-related fields (human, mouse, rat, fly, and C.elegans) for analysis. After sorting the data, 3863561 publications on human genes, 847282 publications on mouse genes, 550865 publications on rat genes, 57346 publications on fly genes, and 12471 publications on \textit{C.elegan}s genes were retained. 

\subsection{Collecting the aging-related and nonaging-related citation numbers on genes}\label{appendix_citation}
Besides the publication numbers on genes, we assumed that the citation information of aging-related publications on a gene was also an indicator describing the research attention paid to such genes. For a publication focusing on aging processes, the prior knowledge from its references assists in shaping and conducting the research. The larger volume of references on one gene suggests a strong influence of this gene on aging-related fields. In this study, we utilized the citation information from iCite \parencite[]{hutchins2019nih} for the analysis of attention to aging-related fields. The data processing protocols are summarized below.

\begin{itemize}
    \item[(1)] In order to count the frequencies that each gene was “cited” by aging-related research for each publication type and species, we first selected aging-related publications both belonging to one publication type and studying the genes of one species. Only genes belonging to the species used for selection in the first step were considered for the following steps. 
    \item[(2)] We secondly extracted the publications cited by the selected publications in the first step.
    \item[(3)] The genes studied by the cited publications in the second step were identified. We supposed these genes were “cited” by the selected aging-related publications. 
    \item[(4)] We counted the frequencies of each gene that appeared in the “cited” genes set. We use these frequencies as the citation numbers on genes for every publication type and species in aging-related research, named aging-related citation numbers.
    \item[(5)] We also calculated the citation numbers in nonaging-related research in a similar way, but selected nonaging-related publications instead of aging-related publications in the first step. 
\end{itemize}

\subsection{Statistically Analyzing the Attentions Across Publication Types by The Random Effect Model}\label{section3.1}
The descriptive quantities including the aging-related and nonaging-related publication numbers for each selected publication type for genes were collected. Here, we assumed that the fraction of aging-related research on genes specific to publication types and species could reflect the attention to aging-related studies. The variation of the fractions among publication types could evaluate their divergence of attention to aging-related studies. The random effect (RE) model was utilized to quantify the fraction of aging-related research considering all publication types and to analyze the source of divergences, more specifically, the research focus on aging processes for a certain species and the distinctiveness of research pattern of publication types. Let $F_{\cdot,t} $ be the estimator of aging-related publication fraction for publication type $t\in T$, say,
\begin{align}
    F_{\cdot,t} = \sum\limits_{g\in G}AP_{g,t}/{\sum\limits_{g\in G}NAP_{g,t}}
\end{align}
where $AP_{g,t}$ and $NAP_{g,t}$ are the aging-related and nonaging-related publication number for gene $g$ and publication type $t$ respectively. $G$ refers to the set containing the total genes with nonzero numbers of publications. The random  effect model can be formulated as:
\begin{align}
    F_{\cdot,t}  = F + \sigma\cdot \varepsilon_t+ h_{t}
\end{align}
    where $F$ is the mean aging-related fraction across pub types, $ \varepsilon_t \sim N(0,1)$ describes the fixed effect, $ h_{t} \sim N(0, \tau^2)$ describes the random effect.
\begin{comment}    
Let $g$ denote the individual gene, $T$ denote the set of all publication types and $t$ denotes the publication type, the RE model can be formulated as:
\begin{align*}
    F_{t}&=\mu_{F}+\sigma_{F,t}\epsilon_{t}+H_{F,t}\\
\epsilon_{t}&\sim N(0,1)\\
H_{F,t}&\sim N(0,\tau_{F}^2)
\end{align*}
where $F_{t}$ is the random variable representing the average fraction of aging-related research on genes for publication type $t\in T$. 
\end{comment}

The RE model accounts for the heterogeneity introduced to $F_{\cdot,t}$ depending on the factors associated with publication types by the heterogeneous variance parameter $\tau^2$, and the sampling uncertainty for the estimates of $F$ by the homogeneous variance parameter $\sigma^2$. We called the R package metafor \parencite[]{viechtbauer2015package} to estimate $\sigma^2$, $\tau^2$ and $F$ after inputting the standard deviations and mean values of fractions of aging-related research for genes of one specific species, i.e.,
\begin{equation*}
    \hat{F} = \frac{\sum\limits_{g}F_{t,g}}{\mbox{total number of genes for publication type $t$}}.
\end{equation*}
Due to the non-normality of the distribution of the fractions for genes, we adopted bootstrap to obtain more precise estimates of standard deviation by sampling from the available fractions for genes with the replacement 1000 times. 

% \bibliographystyle{apacite}
% \bibliography{references}
\printbibliography

\section*{Author Contribution}
This article has been developed collaboratively with both authors, Chenxing Qian and Qingyue Guo. Chenxing Qian: Conceptualization, Data curation, Formal analysis, Investigation, Methodology, Software, Validation, Visualization, Writing—original draft, Writing—review \& editing. Qingyue Guo: Conceptualization, Formal analysis, Investigation, Methodology, Validation, Writing—original draft, Writing—review \& editing.
%\section*{Data Availability}
\section*{Declarations}

Here we declare no conflicts of interest.

\section*{Supplementart Information}
\beginsupplement
\begin{figure}[!htbp]
    \centering
    \includegraphics[width = 1\columnwidth]{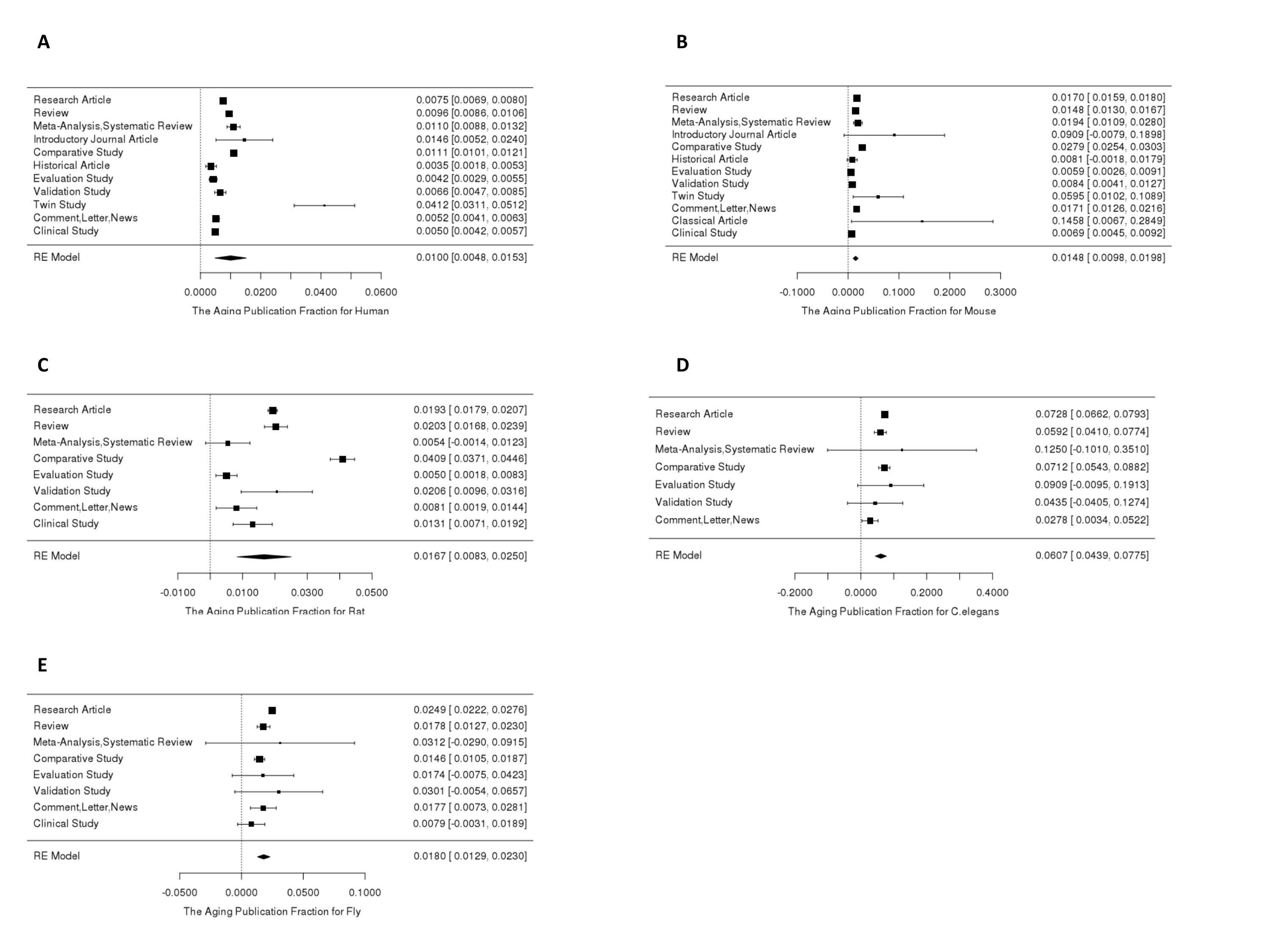}
    \caption{The forest plots displaying the aging-related research publication fraction for various publication types (with the average aging-related research publication fraction regarding all publication types) for human (A), mouse (B), rat (C), \textit{C.elegans} (D) and fly (E).}
    \label{figureS1}
\end{figure}
\begin{figure}[!htbp]
    \centering
     \includegraphics[width = 1\columnwidth]{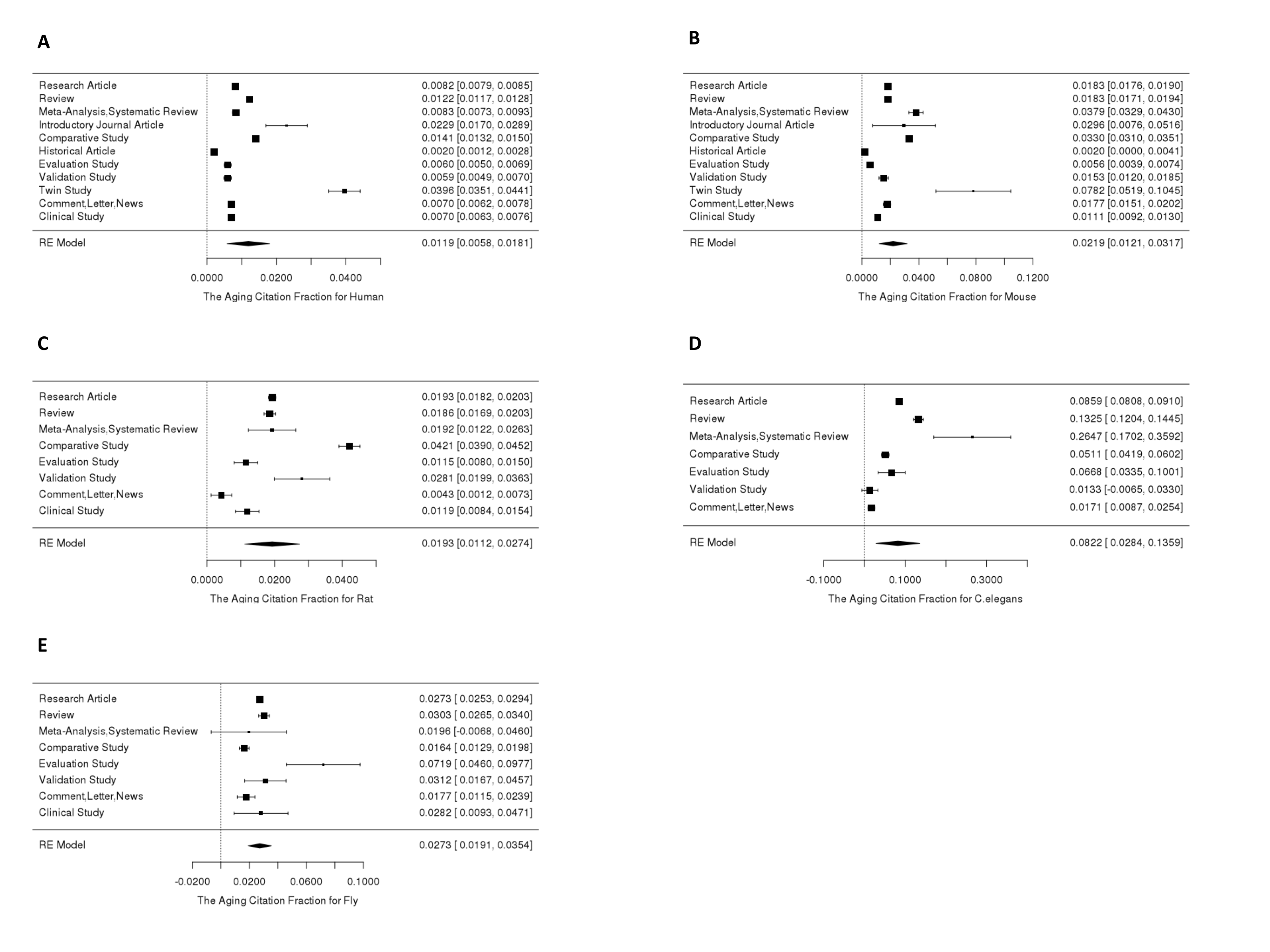}
    \caption{The forest plots displaying the aging-related research citation fraction for various publication types (with the average aging-related research citation fraction regarding all publication types) for human (A), mouse (B), rat (C), \textit{C.elegans} (D) and fly (E).}
    \label{figureS2}
\end{figure}

\begin{table}[!htbp]
    \centering
    \scalebox{0.8}{\begin{tabular}{rrrrrr}
    \hline
    \hline
    \multicolumn{6}{l}{}\\
    \multicolumn{6}{l}{RE Models Diagnostics Using Publication Fractions} \\ 
    \multicolumn{6}{l}{}\\
   \hline
Species & $\tau^2$ & $I^2$ (percentage) & $H^2$ & $Q$ & p-value \\ 
  \hline
Human & 7.4E-05 & 99.529 & 212.270 & 225.222 & 1.7E-04 \\ 
  Mouse & 5.3E-05 & 95.756 & 23.561 & 216.079 & 4.9E-09 \\ 
  Rat & 1.4E-04 & 97.381 & 38.182 & 231.649 & 9.1E-05 \\ 
  C.elegans & 2.4E-04 & 68.097 & 3.134 & 14.198 & 1.4E-12 \\ 
  Fly & 2.5E-05 & 68.672 & 3.192 & 25.175 & 4.0E-12 \\ 
   \hline
   \hline
   \multicolumn{6}{l}{}\\
   \multicolumn{6}{l}{RE Models Diagnostics Using Citation Fractions} \\
   \multicolumn{6}{l}{}\\
   \hline
Species & $\tau^2$ & $I^2$ (percentage) & $H^2$ & $Q$ & p-value \\ 
  \hline
Human & 1.1E-04 & 99.860 & 715.273 & 867.355 & 1.4E-04 \\ 
  Mouse & 2.6E-04 & 99.656 & 290.791 & 740.098 & 1.2E-05 \\ 
  Rat & 1.3E-04 & 98.718 & 78.018 & 339.978 & 3.3E-06 \\ 
  C.elegans & 4.9E-03 & 99.374 & 159.645 & 358.597 & 2.7E-03 \\ 
  Fly & 9.9E-05 & 93.585 & 15.588 & 55.644 & 5.2E-11 \\ 
   \hline
\end{tabular}}
    \label{tab:s1}
    \caption{The key diagnostics of RE model using the publication (and citation) fractions of publication types for 5 species. Tau square refers to the estimated amount of heterogeneity introduced by various publication types. \textit{H square} and \textit{I square} are diagnostics considering the degree of heterogeneity among publication types, among which \textit{I square} refers to the ratio that total estimated heterogeneity over total variability and \textit{H square} refers to the ratio that total variability over sampling variability. Q is the statistics of the heterogeneity test, following the student's distribution with the degree of freedom equals to the number of publication types. p-value shows the significance of heterogeneity.}
    \label{tableS1}
\end{table}

\begin{figure}[!htbp]
    \centering
    \includegraphics[width=1\columnwidth]{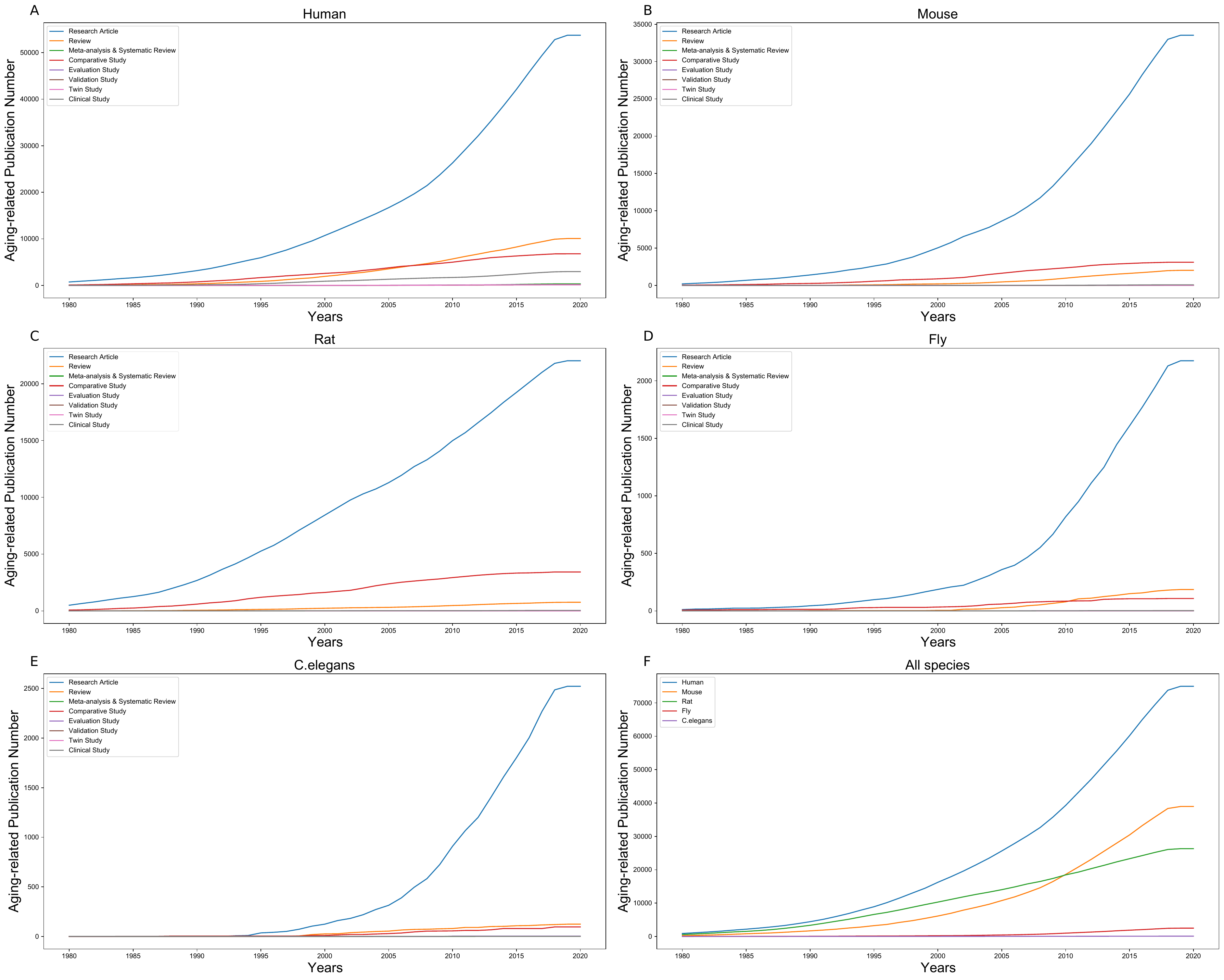}
    \label{fig:accumulated_aging_num}
    \caption{The accumulated aging-related publication numbers of various publication types for 5 species, including Human (A), Mouse (B), Rat (C), Fly (D), and \textit{C. elegans} (E) from 1980 to 2020. The temporal trends of accumulated aging-related publication numbers considering all selected publication types in this study for the five species (F). }
    \label{figureS3}
\end{figure}
\begin{figure}[!htbp]
    \centering
    \includegraphics[width = 1\columnwidth]{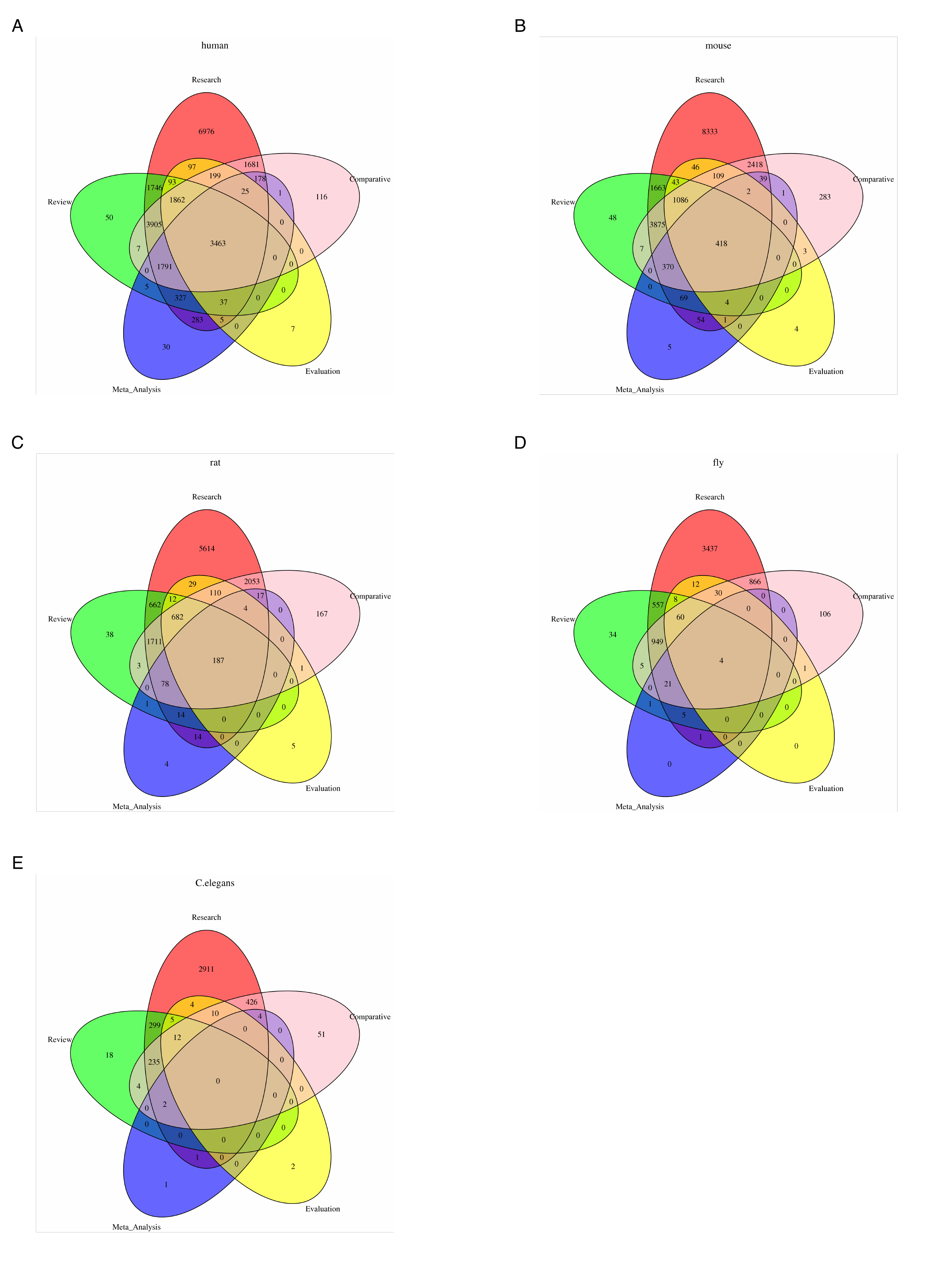}
    \caption{Venn diagrams displaying the scopes, overlaps, and differences of genes studied (genes which own at least one publication in a specific publication type) by various 5 publication types, including Research Article, Review, Meta-Analysis \& Systematic Review (in this figure displayed as Meta\_Analysis), Comparative Study, and Evaluation Study, for Human (A), Mouse (B), Rat (C), Fly (D) and \textit{C.elegans} (E).}
    \label{figureS4}
\end{figure}

\begin{table}[!htbp]
    \centering
    \scalebox{0.8}{\begin{tabular}{rrrrrrrrr}
\hline
\multicolumn{9}{l}{Publication Number Trends for Human} \\
  \hline
 Year & Research & Review & Meta-Analysis, & Comparative & Evaluation & Validation & Twin & Clinical \\
 & Article&&Systematic Review&Study&Study&Study&Study&Study\\
  \hline
2001 & 1129 & 259 & 0 & 146 & 7 & 1 & 4 & 72 \\ 
  2002 & 1172 & 340 & 0 & 154 & 5 & 0 & 2 & 56 \\ 
  2003 & 1183 & 250 & 1 & 321 & 3 & 4 & 2 & 101 \\ 
  2004 & 1198 & 369 & 3 & 265 & 7 & 5 & 1 & 94 \\ 
  2005 & 1301 & 391 & 6 & 314 & 13 & 3 & 6 & 105 \\ 
  2006 & 1416 & 376 & 4 & 310 & 8 & 4 & 6 & 87 \\ 
  2007 & 1560 & 402 & 8 & 185 & 13 & 5 & 10 & 87 \\ 
  2008 & 1765 & 362 & 0 & 215 & 4 & 6 & 4 & 85 \\ 
  2009 & 2299 & 458 & 7 & 230 & 6 & 9 & 6 & 67 \\ 
  2010 & 2559 & 532 & 27 & 265 & 9 & 9 & 5 & 51 \\ 
  2011 & 2889 & 546 & 20 & 336 & 16 & 11 & 4 & 72 \\ 
  2012 & 2917 & 500 & 9 & 290 & 13 & 7 & 33 & 122 \\ 
  2013 & 3186 & 529 & 13 & 343 & 18 & 9 & 5 & 143 \\ 
  2014 & 3335 & 430 & 42 & 185 & 4 & 8 & 2 & 185 \\ 
  2015 & 3465 & 562 & 48 & 182 & 15 & 6 & 5 & 183 \\ 
  2016 & 3720 & 607 & 55 & 161 & 4 & 7 & 13 & 204 \\ 
  2017 & 3587 & 528 & 32 & 148 & 1 & 10 & 0 & 163 \\ 
  2018 & 3395 & 544 & 37 & 140 & 3 & 8 & 2 & 143 \\   
   \hline
   \multicolumn{9}{l}{Publication Number Trends for Mouse} \\
   \hline
 Year & Research & Review & Meta-Analysis, & Comparative & Evaluation & Validation & Twin & Clinical \\
 & Article&&Systematic Review&Study&Study&Study&Study&Study\\
  \hline
2001 & 690 & 18 & 0 & 85 & 0 & 0 & 0 & 4 \\ 
  2002 & 819 & 48 & 0 & 97 & 0 & 0 & 0 & 0 \\ 
  2003 & 604 & 38 & 0 & 197 & 1 & 3 & 0 & 3 \\ 
  2004 & 629 & 70 & 0 & 202 & 3 & 1 & 0 & 0 \\ 
  2005 & 841 & 89 & 1 & 171 & 1 & 1 & 0 & 0 \\ 
  2006 & 835 & 71 & 0 & 174 & 0 & 0 & 1 & 0 \\ 
  2007 & 1075 & 71 & 0 & 174 & 0 & 0 & 0 & 1 \\ 
  2008 & 1208 & 80 & 0 & 117 & 0 & 0 & 0 & 2 \\ 
  2009 & 1568 & 133 & 0 & 134 & 4 & 3 & 0 & 1 \\ 
  2010 & 1851 & 148 & 0 & 124 & 3 & 0 & 1 & 1 \\ 
  2011 & 1925 & 154 & 1 & 139 & 2 & 1 & 0 & 5 \\ 
  2012 & 1929 & 132 & 1 & 179 & 4 & 0 & 1 & 8 \\ 
  2013 & 2166 & 128 & 4 & 121 & 1 & 6 & 0 & 8 \\ 
  2014 & 2225 & 126 & 3 & 78 & 1 & 0 & 0 & 3 \\ 
  2015 & 2238 & 102 & 3 & 71 & 0 & 3 & 1 & 10 \\ 
  2016 & 2609 & 111 & 0 & 68 & 0 & 1 & 1 & 14 \\ 
  2017 & 2405 & 119 & 1 & 46 & 0 & 0 & 0 & 9 \\ 
  2018 & 2326 & 138 & 11 & 44 & 0 & 0 & 0 & 7 \\ 
    \hline
    \multicolumn{9}{l}{Publication Number Trends for \textit{C.elegans}} \\
    \hline
 Year & Research & Review & Meta-Analysis, & Comparative & Evaluation & Validation & Twin & Clinical \\
 & Article&&Systematic Review&Study&Study&Study&Study&Study\\
 \hline
  2001 & 36 & 1 & 0 & 5 & 0 & 0 & 0 & 0 \\ 
  2002 & 22 & 11 & 0 & 5 & 0 & 0 & 0 & 0 \\ 
  2003 & 37 & 6 & 0 & 0 & 0 & 0 & 0 & 0 \\ 
  2004 & 53 & 6 & 0 & 5 & 0 & 1 & 0 & 0 \\ 
  2005 & 42 & 4 & 0 & 4 & 0 & 0 & 0 & 0 \\ 
  2006 & 75 & 11 & 0 & 6 & 0 & 0 & 0 & 0 \\ 
  2007 & 107 & 6 & 0 & 11 & 0 & 0 & 0 & 0 \\ 
  2008 & 87 & 2 & 0 & 8 & 0 & 0 & 0 & 0 \\ 
  2009 & 142 & 4 & 0 & 1 & 0 & 0 & 0 & 0 \\ 
  2010 & 184 & 3 & 0 & 1 & 3 & 0 & 0 & 0 \\ 
  2011 & 158 & 11 & 0 & 5 & 0 & 0 & 0 & 0 \\ 
  2012 & 133 & 0 & 1 & 0 & 0 & 0 & 0 & 0 \\ 
  2013 & 201 & 9 & 0 & 6 & 0 & 0 & 0 & 0 \\ 
  2014 & 208 & 2 & 0 & 13 & 0 & 0 & 0 & 0 \\ 
  2015 & 192 & 6 & 0 & 0 & 0 & 0 & 0 & 0 \\ 
  2016 & 203 & 2 & 0 & 0 & 0 & 0 & 0 & 0 \\ 
  2017 & 264 & 6 & 0 & 0 & 0 & 0 & 0 & 0 \\ 
  2018 & 218 & 4 & 0 & 16 & 0 & 0 & 0 & 0 \\  
   \hline
\end{tabular}}
    \caption{Annual aging-related publication numbers from 2001 to 2018 for Human, Mouse and C.elegans}
    \label{tableS2}
\end{table}

\begin{table}[!htbp]
    \centering
    \scalebox{0.7}{\begin{tabular}{rll}
  \hline
Rank & Research Article & Review \\ 
  \hline
1 & tumor protein p53 & insulin \\ 
  2 & insulin & tumor protein p53 \\ 
  3 & cyclin-dependent kinase inhibitor 2A & insulin-like growth factor 1 \\ 
  4 & apolipoprotein E & sirtuin 1 \\ 
  5 & interleukin 6 & amyloid beta precursor protein \\ 
  6 & tumor necrosis factor & tumor necrosis factor \\ 
  7 & amyloid beta precursor protein & mechanistic target of rapamycin kinase \\ 
  8 & insulin-like growth factor 1 & interleukin 6 \\ 
  9 & cyclin-dependent kinase inhibitor 1A & apolipoprotein E \\ 
  10 & CD4 molecule & growth hormone 1 \\ 
   \hline
 Rank  & Comparative Study & Meta-Analysis,Systematic Review \\ 
  \hline
1 & insulin & apolipoprotein E \\ 
  2 & insulin-like growth factor 1 & insulin \\ 
  3 & interleukin 6 & interleukin 6 \\ 
  4 & apolipoprotein E & C-reactive protein \\ 
  5 & tumor protein p53 & insulin like growth factor 1 \\ 
  6 & tumor necrosis factor & parathyroid hormone \\ 
  7 & growth hormone 1 & interleukin 2 \\ 
  8 & amyloid beta precursor protein & tumor necrosis factor \\ 
  9 & CD4 molecule & CD4 molecule \\ 
  10 & CD8a molecule & interferon alpha 1 \\ 
   \hline
 Rank  & Evaluation Study & Validation Study \\ 
  \hline
1 & insulin & insulin \\ 
  2 & insulin-like growth factor 1 & insulin-like growth factor 1 \\ 
  3 & growth hormone 1 & interleukin 6 \\ 
  4 & tumor protein p53 & C-reactive protein \\ 
  5 & interleukin 6 & anti-Mullerian hormone \\ 
  6 & mechanistic target of rapamycin kinase & adiponectin, C1Q and collagen domain containing \\ 
  7 & apolipoprotein E & apolipoprotein E \\ 
  8 & renin & amyloid beta precursor protein \\ 
  9 & ADAM metallopeptidase with thrombospondin type 1 motif 13 & growth hormone 1 \\ 
  10 & von Willebrand factor & microtubule-associated protein tau \\
   \hline
Rank   & Twin Study & Clinical Study \\ 
  \hline
1 & insulin & insulin \\ 
  2 & apolipoprotein E & insulin-like growth factor 1 \\ 
  3 & C-reactive protein & growth hormone 1 \\ 
  4 & interleukin 6 & interleukin 6 \\ 
  5 & tumor necrosis factor & C-reactive protein \\ 
  6 & E1A binding protein p300 & tumor necrosis factor \\ 
  7 & apolipoprotein A1 & apolipoprotein E \\ 
  8 & kallikrein-related peptidase 3 & sex hormone binding globulin \\ 
  9 & clusterin & growth hormone releasing hormone \\ 
  10 & apolipoprotein B & CD4 molecule \\
   \hline
\end{tabular}}
    \caption{The top 10 human genes that were intensively studied in aging-related research according to the aging-related citation counts on these genes across 8 major publication types}
    \label{tableS3}
\end{table}

\begin{figure}[H]
    \centering
    \includegraphics[width=1\columnwidth]{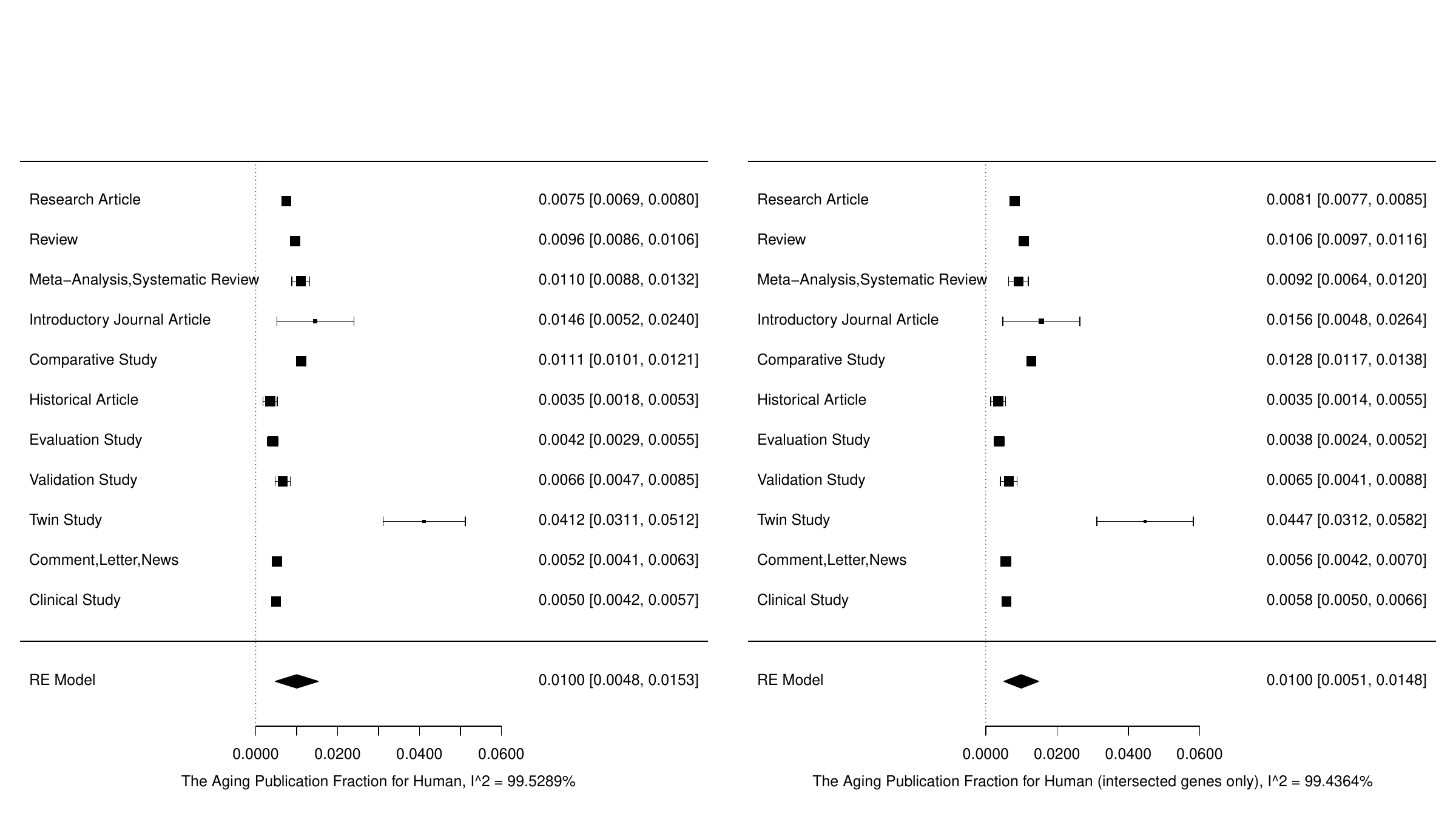}
    \caption{The forest plot forest plots displaying the aging-related research publication fraction across various publication types for human without (left) and with (right) selecting 3463 human genes within the overlap of Research Article, Review, Comparative Study, Meta-Analysis \& Systematic Review and Evaluation Study }
    \label{figureS5}
\end{figure}
\begin{figure}
    \centering
    \includegraphics[width=1\columnwidth]{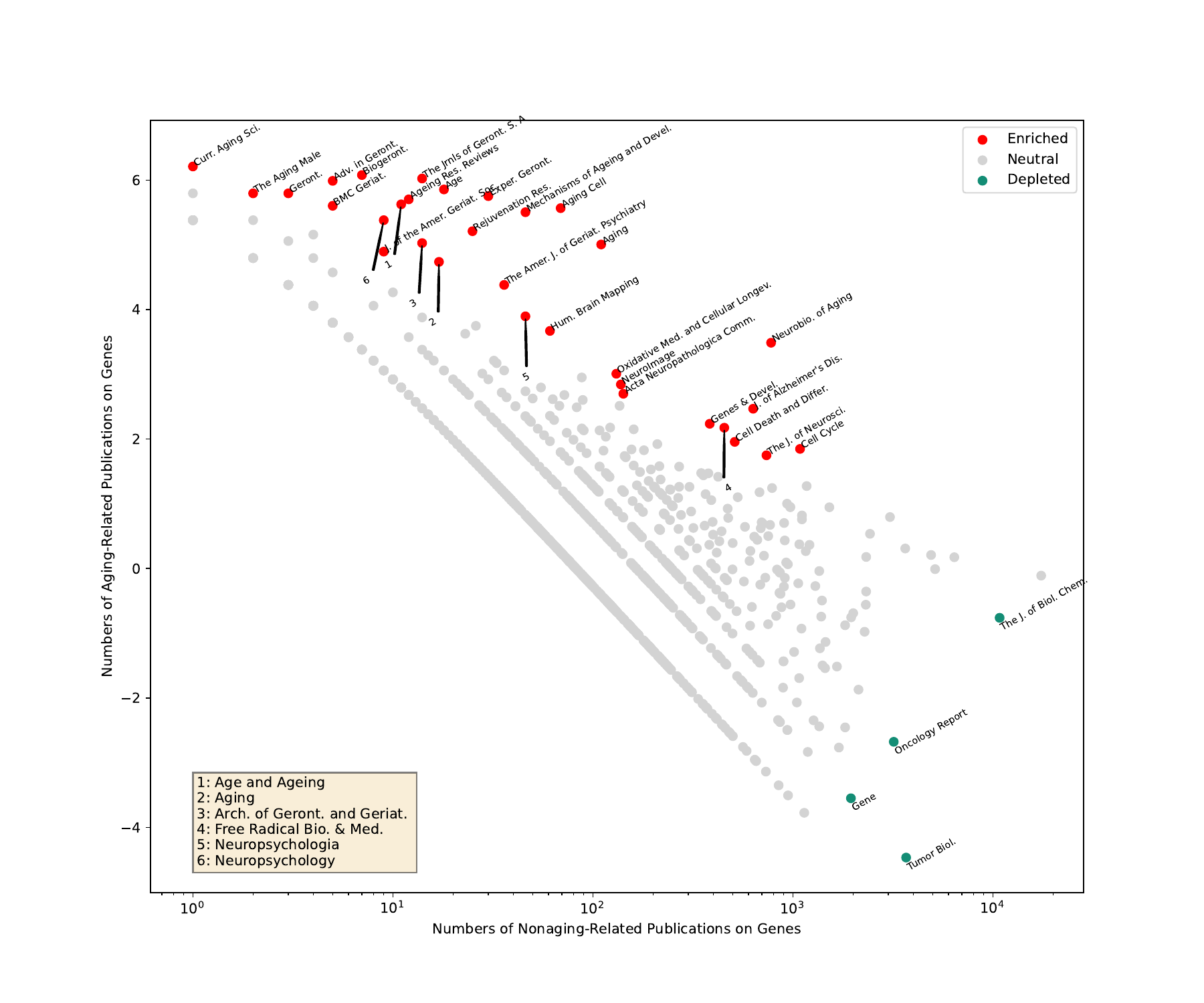}
    \caption{Enrichment (log2 of ratio) for aging-related genetic publications with at least 1 human gene tagged in abstracts or titles over all publications on human genes. Each circle represents an individual journal; red (blue) circles indicate journals that significantly enrich (deplete) aging-related genetic publications, with Bonferroni-corrected p-values using Fisher exact test less than 0.01.
}
    \label{figureS6}
\end{figure}

\begin{table}
\centering
\scalebox{0.7}{\begin{tabular}{rllll}
\hline
Rank &                                             Review &  Bonferroni  &                                  Comparative Study &  Bonferroni  \\
\hline
1 &                            Ageing research reviews &         2.29e-42 &                       Age (Dordrecht, Netherlands) &                    3.89e-36 \\
2 &                           Experimental gerontology &         1.04e-30 &  The journals of gerontology. Series A &                    1.14e-16 \\
3 &               Mechanisms of ageing and development &         7.76e-29 &                           Experimental gerontology &                    1.09e-09 \\
4 &                                        Gerontology &         6.65e-20 &                                              Aging &                    2.31e-07 \\
5 &                                              Aging &         2.58e-12 &                                     Biogerontology &                    2.27e-06 \\
6 &                              Rejuvenation research &         6.05e-09 &                              Neurobiology of aging &                    9.51e-04 \\
7 &                              Current aging science &         1.25e-07 &           Aging clinical and experimental research &                    2.26e-03 \\
8 &                                     Biogerontology &         4.67e-07 &  The journals of gerontology. Series B &                    1.03e-02 \\
9 &                                         Aging cell &         4.21e-06 &                     Australasian J. on ageing &                    4.12e-02 \\
10 &  The journals of gerontology. Series A &         2.49e-05 &                                     Age and ageing &                    4.68e-02 \\
\hline

Rank &                  Meta-Analysis \& Systematic Review &  Bonferroni  &                                   Evaluation Study &  Bonferroni  \\
\hline
1 &                                         Aging cell &                                        0.000120 &                 ACS applied materials \& interfaces &                            1.0 \\
2 &                              Rejuvenation research &                                        0.001198 &                                              Lupus &                            1.0 \\
3 &         Zeitschrift fur Gerontologie und Geriatrie &                                        0.030561 &               Lung cancer (Amsterdam) &                            1.0 \\
4 &  The journals of gerontology. Series A&                                        0.091441 &                                               Lung &                            1.0 \\
5 &                            Ageing research reviews &                                        0.303205 &  Luminescence  &                            1.0 \\
6 &      Medical oncology (Northwood) &                                        1.0 &  Liver transplantation &                            1.0 \\
7 &                                 Medical hypotheses &                                        1.0 &  Liver international  &                            1.0 \\
8 &                                       Medical care &                                        1.0 &                                             Lipids &                            1.0 \\
9 &                          Mediators of inflammation &                                        1.0 &                                      Life sciences &                            1.0 \\
10 &  Medical science monitor  &                                        1.0 &                          Macromolecular bioscience &                            1.0 \\
\hline
Rank &                                   Validation Study &  Bonferroni  &                                         Twin Study &  Bonferroni  \\
\hline
1 &                               ACS chemical biology &                            1.0 &  The journals of gerontology. Series A&                 0.47 \\
2 &                     J. of virological methods &                            1.0 &                         Acta dermato-venereologica &                 1.0 \\
3 &                                J. of virology &                            1.0 &              Korean J. of ophthalmology : KJO &                 1.0 \\
4 &  J. of voice &                            1.0 &                                           Leukemia &                 1.0 \\
5 &                               Kidney international &                            1.0 &                                Leukemia \& lymphoma &                 1.0 \\
6 &  KSSTA &                            1.0 &           Metabolic syndrome and related disorders &                 1.0 \\
7 &                             La Medicina del lavoro &                            1.0 &              Metabolism: clinical and experimental &                 1.0 \\
8 &  Laboratory investigation &                            1.0 &                               Molecular metabolism &                 1.0 \\
9 &                           Lancet (London) &                            1.0 &                               Molecular psychiatry &                 1.0 \\
10 &       Learning \& memory (NY) &                            1.0 &                                             Nature &                 1.0 \\

\hline
Rank &                                     Clinical Study &  Bonferroni  &                Comment, News, Letter &  Bonferroni\\
\hline
1 &               Mechanisms of ageing and development &                 3.33e-23 &                                Aging &                        1.87e-09 \\
2 &                           Experimental gerontology &                 7.27e-12 &        Cell cycle (Georgetown) &                        1.13e-05 \\
3 &                       Age (Dordrecht, Netherlands) &                 1.72e-06 &                            Neurology &                        1.28e-02 \\
4 &  The journals of gerontology. Series A. &                 1.00e-05 &      Heart (British Cardiac Society) &                        2.27e-02 \\
5 &           Aging clinical and experimental research &                 5.96e-05 &    J. of psychosomatic research &                        7.99e-02 \\
6 &  J. of bone and mineral research &                 3.88e-02 &                Rejuvenation research &                        2.38e-01 \\
7 &  JIFR &                 4.27e-02 &  Human vaccines \& immunotherapeutics &                        7.89e-01 \\
8 &                                     BMC geriatrics &                 4.27e-02 &                       Cell stem cell &                        9.24e-01 \\
9 &                        Experimental aging research &                 7.71e-02 &                      Nature medicine &                        1\\
10 &           The J. of nutrition, health \& aging &                 1.27e-01 &                    Nature metabolism &                        1 \\
\hline
\label{tab:Jrnl2}
\end{tabular}}
\caption{The top 10 journals with the highest proportion of aging-related publications among 8 different publication types, ranked by Bonferroni corrected p-values (significance). JIFR: The J. of Injury, Function, and Rehabilitation; KSSTA:  Knee Surgery, Sports Traumatology, Arthroscopyt; }
\label{tableS4}
\end{table}

\end{document}